\documentclass[12pt, draftclsnofoot, journal, letterpaper, onecolumn]{IEEEtran}

\IEEEoverridecommandlockouts
\usepackage{amsfonts}
\usepackage[dvips]{graphicx}
\usepackage{times}
\usepackage{cite}
\usepackage{amsmath}
\usepackage{cases}
\usepackage{array}
\usepackage{dsfont}
\usepackage{amssymb}

\usepackage{stfloats}
\usepackage{slashbox}
\usepackage{graphicx}
\usepackage{footnote}
\usepackage{booktabs}
\usepackage{array}
\usepackage{multirow}
\usepackage{bm}
\usepackage{empheq}
\usepackage[labelformat=simple]{subcaption}

\usepackage{amsthm}
\usepackage{algorithm}
\usepackage{algorithmic}
\usepackage{color}
\usepackage{diagbox}
\theoremstyle{plain}
\newtheorem{theorem}{Theorem}
\newtheorem{corollary}{Corollary}
\newtheorem{lemma}{Lemma}

\theoremstyle{definition}

\newtheorem{remark}{Remark}
\begin{document}

\abovedisplayskip=2pt 
\belowdisplayskip=2pt 

\title{Heterogeneous Coded Distributed Computing: Joint Design of File Allocation and Function Assignment}
\author{\IEEEauthorblockN{Fan Xu and Meixia Tao}\\
\IEEEauthorblockA{Department of Electronic Engineering, Shanghai Jiao Tong University, Shanghai, China \\Emails: \{xxiaof, mxtao\}@sjtu.edu.cn}\\
\thanks{This work is supported by the National Natural Science Foundation of China under grant 61571299 and the Shanghai Key Laboratory Funding under grant STCSM18DZ2270700.}}
\maketitle

\begin{abstract}
This paper studies the computation-communication tradeoff in a heterogeneous MapReduce computing system where each distributed node is equipped with different computation capability. We first obtain an achievable communication load for any given computation load and any given function assignment at each node. The proposed file allocation strategy has two steps: first, the input files are partitioned into disjoint batches, each with possibly different size and computed by a distinct node; then, each node computes additional files from its non-computed files according to its redundant computation capability. In the Shuffle phase, coded multicasting opportunities are exploited thanks to the repetitive file allocation among different nodes. Based on this scheme, we further propose the computation-aware and the shuffle-aware function assignments. We prove that, by using proper function assignments, our achievable communication load for any given computation load is within a constant multiplicative gap to the optimum in an equivalent homogeneous system with the same average computation load. Numerical results show that our scheme with shuffle-aware function assignment achieves better computation-communication tradeoff than existing works in some cases.
\end{abstract}

\section{Introduction}\label{sec introduction}
Driven by the rapid growth of machine learning and data science,  modern computation paradigm has shifted from conventional one-processor systems towards large-scale distributed computing systems, such as Hadoop. MapReduce is a prevalent framework for distributed computing \cite{MapReduceOrigin}, where the computation is decomposed into ``Map'' and ``Reduce'' stages. Each distributed node first computes the Map functions from its locally available files  to generate intermediate values (IVs). These IVs are then shuffled among nodes, so that each node can use these IVs to compute the Reduce functions, and obtain the final output values. In this framework, data shuffling among distributed nodes appears to be a major bottleneck of the distributed computing systems due to the large communication load. For example, $70\%$ of the execution time is for data shuffling when running ``SelfJoin'' on the Amazon EC2 cluster \cite{EC2}.

To alleviate the communication bottleneck, \emph{coded distributed computing} is proposed by \cite{LiSongZeMapReduce} in the MapReduce framework. It leverages the redundant computation capabilities at nodes by carefully designing input file allocation in the Map phase so as to exploit coded multicasting opportunities and hence reduce communication load in the Shuffle phase. The optimal tradeoff between the computation load in the Map phase and the communication load in the Shuffle phase  is derived in \cite{LiSongZeMapReduce}, which finds that increasing computation load of the Map phase by $r$ can reduce communication load of the Shuffle phase by the same factor $r$. This  idea of coded distributed computing has since been extended widely, e.g., \cite{JimingyueNewCombinatorial,LeveragingCoding,Yanqifa3C,Loadscheduling,LisongzeScalable,ChenjinyuanWirelessMR}. In particular, \cite{JimingyueNewCombinatorial,LeveragingCoding} propose new coded distributed computing schemes, \cite{Yanqifa3C} studies distributed computing with storage constraints at nodes, \cite{Loadscheduling} studies distributed computing under time-varying excess computing resources, and  \cite{LisongzeScalable,ChenjinyuanWirelessMR} studies the wireless distributed computing systems.

Note that all these works \cite{LiSongZeMapReduce,JimingyueNewCombinatorial,LeveragingCoding,Loadscheduling,Yanqifa3C,LisongzeScalable,ChenjinyuanWirelessMR} focus on homogeneous computing systems where each node is allocated the same number of input files and assigned the same number of output functions. In practice, however, nodes are equipped with different storage and computation capabilities. The authors in \cite{Wangchengwei} propose a scheme for the heterogeneous system with $K=3$ nodes, which achieves the optimal computation-communication tradeoff, and present an algorithm to generalize this scheme to the system with $K>3$ nodes. The authors in \cite{edge-facilitated} extend \cite{Wangchengwei} to a wireless heterogeneous distributed computing system with $K=3$ nodes, where each node is connected with others through a common access point, and obtain an achievable communication load region and a converse for the uplink-downlink transmission pair. The authors in \cite{Chenjinyuan} study the minimum computation load given heterogeneous communication constraints, and characterize the optimal computation load for the system with $K=2$ or $3$ nodes, and for the system with $K>3$ nodes in certain cases. These works \cite{Wangchengwei,Chenjinyuan,edge-facilitated} reveal that, in a heterogeneous system, the file allocation over nodes is non-cyclically symmetric and should be carefully designed so that coded multicasting opportunities are created as many as possible in the Shuffle phase to obtain the optimal computation-communication tradeoff. However, they only consider heterogeneous file allocation in the Map phase due to different storage size, and still assume homogeneous function assignment in the Reduce phase without taking the different  computation capabilities across nodes into account. The authors in \cite{JimingyueCascade,JimingyueNoncascade} consider the heterogeneous systems where each node is assigned different number of output functions. Both works obtain an achievable communication load which is within a constant multiplicative gap to the optimum given the considered function assignment. They find that, by assigning more output functions to nodes with more input files, their proposed schemes even outperform the optimal scheme in an equivalent homogeneous system \cite{LiSongZeMapReduce} in some cases. However, the heterogeneous systems considered in \cite{JimingyueCascade,JimingyueNoncascade} consist of multiple homogeneous systems where nodes in each system have the same storage and computation capabilities but differ from nodes in other systems, and is thus not suitable to general heterogeneous systems.

In this paper, we study the computation-communication tradeoff in a general heterogeneous MapReduce computing system. The system consists of $K$ nodes, where each node $k$ computes the map functions of $m_kN$ files from the total $N$ input files, and $m_k$ is known as its computation load. We first obtain an achievable communication load in a closed-form expression for any given computation load and any given function assignment at each node. The proposed file allocation strategy in the Map phase has two steps: first, the $N$ input files are partitioned into $K$ disjoint batches with possibly different sizes, each computed by a distinct node and referred to as its compulsory files; then, each node further computes the compulsory files of other nodes according to its redundant computation capability, and we refer to these files as its optional files. In the proposed data shuffling strategy, each node distributes the IVs computed from its compulsory files to the requiring nodes. Given the repetitive file allocation arising from the design of optional files, coded multicasting opportunities are exploited, where zero-padding is used to generate the coded messages. We then propose two function assignments to further reduce the communication load. In the computation-aware function assignment, the number of output functions assigned to each node is proportional to its computation load. In the shuffle-aware function assignment, all the output functions are properly assigned to nodes with high computation load to reduce traffic load in the Shuffle phase. The achievable communication load obtained by these two function assignment methods is proved to be within a constant multiplicative gap to the optimal load $L_{\textrm{Hom}}^*$ in an equivalent homogeneous system with the same average computation load \cite{LiSongZeMapReduce}. Numerical results show that the communication load with shuffle-aware function assignment is smaller than $L_{\textrm{Hom}}^*$  and achievable loads in other works in some cases.

\textbf{Notations:} For $K\in\mathbb{Z}^+$, $[K]$ denotes the set $\{1,2,\ldots,K\}$. For $a\!<\!b$, $a,b\in\mathbb{Z}$, $[a\!:\!b]$ denotes the set $\{a,a+1,\ldots,b-1,b\}$. $[a]^{1\times K}$ denotes the $1\!\times\! K$ vector with all entries being $a$.

\section{System Model}\label{sec system}
We consider a distributed computing system which aims to compute $Q$ output functions from $N$ input files using $K$ distributed nodes, for some positive integers $Q$, $N$, and $K$. The input files are denoted by $\{f_1,\ldots,f_N\}$, each of size $F$ bits, and the output functions are denoted by $\{\phi_1,\ldots,\phi_Q\}$, where $\phi_q$, for $q\in[Q]$, maps all the input files into the output value $u_q=\phi_q(f_1,\ldots,f_N)\in \mathbb{F}_{2^B}$ with length $B$ bits. The computing system follows the MapReduce framework as in \cite{MapReduceOrigin,LiSongZeMapReduce}, where the computation of each output function can be decomposed as
\begin{align}
\phi_q(f_1,\ldots,f_N)=h_q(g_{q,1}(f_1),\ldots,g_{q,N}(f_N)).\label{eqn mapreduce}
\end{align}
Here, $\mathbf{g}_{n}=(g_{1,n},\ldots,g_{Q,n})$ is the Map function that maps input file $f_n$ into $Q$ IVs $v_{q,n}=g_{q,n}(f_n)\in\mathbb{F}_{2^T}$, for $q\in[Q]$, each with length $T$ bits; and $h_q$ is the Reduce function that maps the IVs of the output function $\phi_q$ in all input files into the output value $u_q=h_q(v_{q,1},\ldots,v_{q,N})$. Following this decomposition, the MapReduce computing system consists of three phases: \emph{Map}, \emph{Shuffle} and \emph{Reduce}.


\textbf{Map phase}: Each node $k$, for $k\in[K]$, stores $M_k$ files from the $N$ input files, denoted by $\mathcal{M}_k\subset\{f_1,\ldots,f_N\}$ with $|\mathcal{M}_k|=M_k< N$. It computes the Map function of each file $f_n\in\mathcal{M}_k$ to obtain the IVs $\{v_{q,n}:q\in[Q],f_n\in\mathcal{M}_k\}$. We assume that $\sum_{k\in[K]}M_k\ge N$ so that the Map function of each file can be computed at least once. Define the computation load of node $k$, denoted by $m_k$, as the number of Map functions it computes normalized by the  total number of input files $N$, i.e., $m_k\triangleq\frac{M_k}{N}$, and $\mathbf{m}\triangleq[m_1,\ldots,m_K]$ as the overall computation load vector. We have $\sum_{k\in[K]}m_k\ge1$, and $m_k< 1,\forall k\in[K]$. Without loss of generality, we assume that $m_1\le m_2\le\cdots\le m_K$.

Each node $k$ is assigned to compute a subset of $W_k$ output functions from the total $Q$ functions, denoted by $\mathcal{W}_k\subseteq\{\phi_1,\ldots,\phi_{Q}\}$ with $|\mathcal{W}_k|=W_k$. Note that, unlike \cite{LiSongZeMapReduce,JimingyueNewCombinatorial,Wangchengwei,edge-facilitated,LisongzeScalable,ChenjinyuanWirelessMR,Yanqifa3C,LeveragingCoding,Chenjinyuan}, $W_k$ may vary for different $k$. Similar to \cite{JimingyueNoncascade,Wangchengwei,LisongzeScalable,edge-facilitated,ChenjinyuanWirelessMR,Yanqifa3C,LeveragingCoding,Chenjinyuan}, we assume that $\mathcal{W}_j\cap \mathcal{W}_k=\emptyset$ for $j\neq k$ so that each function is assigned to exactly one node. Thus, we have $\sum_{k\in[K]}W_k=Q$. Define the function assignment of node $k$, denoted by $w_k$, as the number of output functions it computes normalized by the total number of output functions $Q$, i.e., $w_k=\frac{W_k}{Q}$, and $\mathbf{w}\triangleq[w_1,\ldots,w_K]$ as the overall function assignment vector. Then, we have $\sum_{k\in[K]}w_k=1$.

\textbf{Shuffle phase}: To compute the assigned output functions, each node needs the IVs which are not computed locally in the Map phase. Thus, each node $k$ creates a message $X_k\in \mathbb{F}_{2^{\ell_k}}$ with length $\ell_k$ bits as a function of the IVs computed locally in the Map phase, i.e., $X_k=\psi_k(\{v_{q,n}:q\in[Q],f_n\in{\mathcal{M}_k}\})$ for some encoding function $\psi_k$, and broadcasts it to the rest nodes. Similar to \cite{LiSongZeMapReduce}, the communication load $L$ is defined as $L\triangleq\frac{\sum_{k\in[K]}\ell_k}{QNT}$, which characterizes the normalized total number of bits communicated among the $K$ nodes in the Shuffle phase.

\textbf{Reduce phase}: Each node $k$ uses its local results $\{v_{q,n}:q\!\in\![Q],f_n\!\in\!\mathcal{M}_k\}$ computed in the Map phase and the messages $\{X_1,\ldots,X_K\}$ communicated in the Shuffle phase to construct the IVs $\{v_{q,n}:\phi_q\!\in\! \mathcal{W}_k,n\!\in\![N]\}$, and then computes the Reduce functions of its assigned output functions $\mathcal{W}_k$.

For a given computation load $\mathbf{m}$ and a given function assignment $\mathbf{w}$, the minimum communication load in the Shuffle phase is defined as $L^*(\mathbf{m},\mathbf{w})$. In this paper, we aim to jointly design file allocation and function assignment $\mathbf{w}$ for any given computation load $\mathbf{m}$ in the Map phase, so as to minimize the communication load in the Shuffle phase. For ease of analysis, we assume that $Q$ and $N$ are sufficiently large to ensure that the number of files and the number of functions assigned to each node are integers by our scheme.

\section{Achievable Communication Load at Given Function Assignment $\mathbf{w}$}\label{sec achievable scheme heterogeneous}
In this section, we present our achievable scheme in the MapReduce computing system for arbitrary computation load $\mathbf{m}$ and arbitrary function assignment $\mathbf{w}$, and obtain the achievable communication load $L_A(\mathbf{m},\mathbf{w})$  at a given computation load $\mathbf{m}$ and function assignment $\mathbf{w}$. The design of specific function assignment $\mathbf{w}$ to minimize the communication load  shall be presented in the next section.

We first present our scheme through a 4-node example, and then proceed to the general scheme.

\subsection{An Example}\label{sec example}
In this subection, we use a 4-node MapReduce computing system with $\mathbf{m}=[\frac{1}{5},\frac{1}{3},\frac{1}{3},\frac{1}{2}]$ and $\mathbf{w}=[\frac{1}{8},\frac{1}{4},\frac{1}{6},\frac{11}{24}]$ as an example to illustrate the proposed scheme.

\subsubsection{Map phase design}
The proposed file allocation strategy in the Map phase has two steps. In the first step, the strategy is to allocate the input files among all the nodes exclusively as equal as possible, which may result in that the nodes with low computation load are exhausted while the nodes with high computation load still have extra computing capacity. Specifically, in this example, we first allocate a batch of $\frac{1}{5}N$ files to node 1 to fill its computation load  since $m_1=\frac{1}{5}<\frac{1}{4}$, then we equally divide the rest $\frac{4}{5}N$ files into three batches, each with size $\frac{4}{15}N$, and allocate them to the other three nodes. Denote the disjoint file batch allocated to node $k$ using this strategy as $\mathcal{N}_k$, with size $l_k$. In this example, we have $l_1=\frac{1}{5}$ and $l_2=l_3=l_4=\frac{4}{15}$.

\begin{figure}[tbp]
\centering
\includegraphics[scale=0.5]{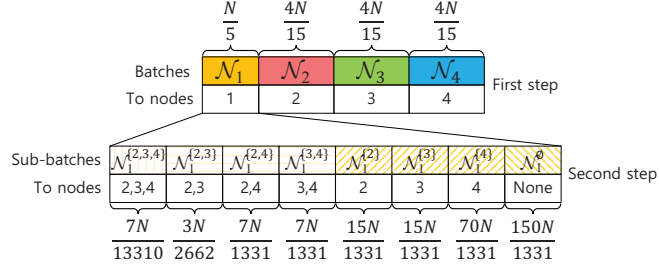}
\vspace{-5pt}
\caption{The two-step file allocation strategy.}\label{Fig example map}
\vspace{-22pt}
\end{figure}

After the first step, each node $k$ is still able to compute the Map functions of $(m_k-l_k)N$ more files from the remaining $(1-l_k)N$ files $\{f_1\ldots,f_N\}\setminus\mathcal{N}_k$. Define $P_k\!\triangleq\!\frac{m_k-l_k}{1-l_k}$ as the surplus computation ratio of node $k$. In this example, we have $[P_1,P_2,P_3,P_4]\!=\![0,\frac{1}{11},\frac{1}{11},\frac{7}{22}]$. In the second step, we further partition each batch $\mathcal{N}_k$, for $k\in[4]$, into $8$ sub-batches $\{\mathcal{N}_k^{\Psi}:\Psi\subseteq[4]\setminus\{k\}\}$. Each sub-batch $\mathcal{N}_k^{\Psi}$ has $l_k^{\Psi}N$ files with
\begin{align}
  l_k^{\Psi}\triangleq l_k\prod_{i\in\Psi}P_i\prod_{i\in[4]\setminus \{\Psi,k\}}(1-P_i),\label{eqn example file size}
\end{align}
and is further allocated to nodes in set $\Psi\subseteq[4]\setminus\{k\}$. For example, $\mathcal{N}_1^{\{2,3\}}$ is the sub-batch of $\mathcal{N}_1$ which is allocated exclusively to node $k=1$ in the first step then re-allocated to nodes $\Psi= \{2,3\}$ in the second step. There are $l_1^{\{2,3\}}N\!=\!l_1P_2P_3(1-P_4)N\!=\!\frac{3}{2662}N$ files in $\mathcal{N}_1^{\{2,3\}}$. Note that the file allocation strategy in the second step is inspired by the decentralized cache placement \cite{decentralized,Wangsinong}, since the number of files exclusively computed by nodes $\Psi\cup\{k\}$ in $\mathcal{N}_k$ can be viewed as the number of file bits exclusively cached by these nodes by decentralized cache placement which converges to $l_k^{\Psi}N$ with high probability for a sufficiently large file number $N$. Fig. \ref{Fig example map} shows our two-step file allocation strategy, where the second step is illustrated by using $\mathcal{N}_1$ as an example.

By using the proposed two-step file allocation strategy, it is easy to verify that each node $k$ computes the Map functions of $m_kN$ files. For each node $k$, the files selected in the first step, i.e., $\mathcal{N}_k$, are referred to as its \emph{compulsory} files, while the files selected in the second step, denoted by $\mathcal{O}_k$, are referred to as its \emph{optional} files. We have $\mathcal{M}_k=\mathcal{N}_k\cup\mathcal{O}_k$. Note that node 1 has no optional file, i.e., $\mathcal{O}_1=\emptyset$, since $m_1=l_1$. We refer to the nodes that have no optional file as \emph{low-computation-load} (LowCL) nodes, and refer to the rest nodes as \emph{high-computation-load} (HighCL) nodes. Here, node 1 is the LowCL node, and nodes $[2\!:\!4]$ are the HighCL nodes.

\subsubsection{Shuffle phase design}
After the Map phase, each node $k$, for $k\!\in\![4]$, needs the IVs of the other $(1-m_k)N$ files to compute the Reduce functions of its assigned functions $\mathcal{W}_k$. Meanwhile, it should distribute the IVs computed from its compulsory files $\mathcal{N}_k$ to the requiring nodes. We use unicasting to deliver the IVs needed by LowCL nodes, i.e., node 1, and use coded multicasting to deliver the IVs needed by HighCL nodes, i.e., nodes $[2:4]$.

\textbf{Communication to node 1}: Since node 1 has no optional file, each node $k$, for $k\!\in[2\!:\!4]$, directly unicasts the IVs of the output functions $\mathcal{W}_1$ in its compulsory files $\mathcal{N}_k$ to node 1, given by
\begin{align}
\mathcal{V}_{k\rightarrow\{1\}}=\left\{v_{q,n}:\phi_q\in \mathcal{W}_{1}, f_n\in\mathcal{N}_k\right\}.\notag
\end{align}
Since $|\mathcal{W}_1|\!=\!w_1Q\!=\!\frac{1}{8}Q$ and $|\mathcal{N}_k|\!=\!l_kN\!=\!\frac{4}{15}N$, for $k\!\in\![2\!:\!4]$, the communication load from node $k$ to node 1 is given by
\begin{align}
  L_{k\rightarrow\{1\}}=\frac{w_1Q\cdot l_kN\cdot T}{QNT}=w_1l_k=\frac{1}{30}.\notag
\end{align}
Combining all three nodes $[2:4]$, the total communication load to node 1 is given by
\begin{align}
  L_1=\sum_{k=2}^4L_{k\rightarrow\{1\}}=w_1\sum_{k=2}^4l_k=w_1(1-l_1)=\frac{1}{10}.\label{eqn example node 1 needed load}
\end{align}

After receiving its desired IVs computed from files $\mathcal{N}_2\cup\mathcal{N}_3\cup\mathcal{N}_4$, and combining the IVs computed locally from files $\mathcal{N}_1$, node 1 can successfully compute the Reduce functions of its assigned functions $\mathcal{W}_1$ and obtain the output values.

\textbf{Communication to nodes $[2:4]$}: Since each node $k\in[2:4]$ already has the IVs of the assigned functions $\mathcal{W}_k$ in its optional files $\mathcal{O}_k$, each node $i\in[4]\setminus \{k\}$ only needs to send the IVs of the functions $\mathcal{W}_k$ in the rest of its compulsory files $\mathcal{N}_i$ to node $k$, given by
\begin{align}
  \left\{v_{q,n}:\phi_q\in\mathcal{W}_k,f_n\in\mathcal{N}_i\setminus\mathcal{O}_k\right\}.\notag
\end{align}
Since the Map function of each file in $\mathcal{N}_i\!\setminus\!\mathcal{O}_k$ is possibly computed by node $j\!\in[2\!:\!4]\!\setminus\!\{k,i\}$ given the second step of file allocation in the Map phase, coded multicasting opportunities can be exploited.

We first take the communication from node 1 to node set $\{2,3\}$ as an example. The IVs of functions $\mathcal{W}_2$ in files $\mathcal{N}_1^{\{3\}}$ are needed by node 2 and available at node 3, given by
\begin{align}
  \mathcal{V}_{1\rightarrow2}^{\{3\}}=\{v_{q,n}:\phi_q\in \mathcal{W}_2, f_n\in\mathcal{N}_1^{\{3\}}\}.\notag
\end{align}
There are $w_2Q\cdot l_1^{\{3\}}N=w_2Q\cdot l_1\cdot P_3(1-P_2)(1-P_4)N=\frac{15}{5324}QN$ IVs in $\mathcal{V}_{1\rightarrow2}^{\{3\}}$. On the other hand, the IVs of functions $\mathcal{W}_3$ in files $\mathcal{N}_1^{\{2\}}$ are needed by node 3 and available at node 2, given by
\begin{align}
  \mathcal{V}_{1\rightarrow3}^{\{2\}}=\{v_{q,n}:\phi_q\in \mathcal{W}_3, f_n\in\mathcal{N}_1^{\{2\}}\}.\notag
\end{align}
There are $w_3Q\cdot l_1^{\{2\}}N=w_3Q\cdot l_1\cdot P_2(1\!-\!P_3)(1\!-\!P_4)N=\frac{5}{2662}QN$ IVs in $\mathcal{V}_{1\rightarrow3}^{\{2\}}$. Padding $\frac{5}{5324}QNT(=\!\frac{15}{5324}QNT\!-\!\frac{5}{2662}QNT)$ zero bits to IV set $\mathcal{V}_{1\rightarrow3}^{\{2\}}$, and combining the IVs in  $\mathcal{V}_{1\rightarrow2}^{\{3\}}$ and $\mathcal{V}_{1\rightarrow3}^{\{2\}}$ using bit-wise XOR, node 1 can send
\begin{align}
  \mathcal{V}_{1\rightarrow\{2,3\}}=\mathcal{V}_{1\rightarrow2}^{\{3\}}\oplus \mathcal{V}_{1\rightarrow3}^{\{2\}}\notag
\end{align}
to nodes $\{2,3\}$, where $\oplus$ denotes the bit-wise XOR operation. After receiving $\mathcal{V}_{1\rightarrow\{2,3\}}$, both node 2 and 3 can obtain their desired IVs. The communication load from node 1 to node set $\{2,3\}$ is given by
\begin{align}
  L_{1\!\rightarrow\!\{2,3\}}\!=\!\frac{\max\left\{w_2Q\cdot l_1^{\{3\}}NT,w_3Q\cdot l_1^{\{2\}}NT\right\}}{QNT}=\!\frac{15}{5324}.\notag
\end{align}

In general, by using zero-padding and XOR combining, the coded message multicasted from an arbitrary node $k$ to an arbitrary node set $\Psi\subseteq[2:4]\setminus\{k\}$ is given by
\begin{align}
  \mathcal{V}_{k\rightarrow\Psi}\triangleq\bigoplus_{i\in\Psi}\mathcal{V}_{k\rightarrow i}^{\Psi\setminus\{i\}}\label{eqn example xor}
\end{align}
where $\mathcal{V}_{k\rightarrow i}^{\Psi\setminus\{i\}}\triangleq\{v_{q,n}:\phi_q\in \mathcal{W}_i, f_n\in\mathcal{N}_k^{\Psi\setminus\{i\}}\}$ are the IVs needed by node $i\in\Psi$ and available at nodes $\Psi\setminus\{i\}$. The communication load is given by
\begin{align}
  L_{k\rightarrow\Psi}\!=&\frac{\max_{i\in\Psi}|\mathcal{V}_{k\rightarrow i}^{\Psi\setminus\{i\}}|}{QNT}\notag\\
  =&\frac{\max\limits_{i\in\Psi}\!\Big\{\!l_k\!\prod\limits_{j\in\Psi}\!P_j\!\cdot\!\!\prod\limits_{j\in[2:4]\setminus\Psi,j\neq k}\!\!(1\!-\!P_j)\!\cdot\!\frac{w_i(1\!-\!P_i)}{P_i}QNT\!\Big\}}{QNT}\notag\\
  =&l_k\!\prod_{j\in\Psi}\!P_j\!\cdot\!\!\!\prod_{j\in[2:4]\setminus\Psi,j\neq k}\!\!(1\!-\!P_j)\!\cdot\! \max_{i\in\Psi}\!\left\{\!\frac{w_i(1\!\!-\!\!P_i)}{P_i}\!\right\}\!.\label{eqn example one set load}
\end{align}

Summing up the load from each node $k$ to each set $\Psi\!\subseteq\![2\!:\!4]\!\setminus\!\{k\}$, the total load to nodes $[2\!:\!4]$ is given by
\begin{align}
  L_{[2:4]}
  =&w_2(1\!\!-\!\!P_2)(l_1\!+\!l_3\!+\!l_4)\!+\!w_3(1\!\!-\!\!P_2)(1\!\!-\!\!P_3)(l_1\!\!+\!\!\frac{l_2}{1\!-\!P_2}\!\!+\!l_4)\notag\\
  &+\!w_4(1\!\!-\!\!P_2)(1\!\!-\!\!P_3)(1\!\!-\!\!P_4)(l_1\!\!+\!\!\frac{l_2}{1\!-\!P_2}\!\!+\!\!\frac{l_3}{1\!-\!P_3})\notag\\
  =&\frac{689}{1452}.\label{eqn example node 234 needed load}
\end{align}

Based on this shuffle strategy, each node $i\!\in\![2\!:\!4]$ can obtain its needed IVs $\mathcal{V}_{k\rightarrow i}^{\Psi}$ from coded message $\mathcal{V}_{k\rightarrow\Psi\cup\{i\}}$, for $k\in[4]\setminus\{i\},\Psi\!\subseteq\![2\!:\!4]\!\setminus\!\{i,\!k\}$. Combining the IVs computed locally from files $\mathcal{M}_i$, node $i$ can successfully compute the Reduce functions of its assigned functions $\mathcal{W}_i$ and obtain the output values.

Summing up \eqref{eqn example node 1 needed load} and \eqref{eqn example node 234 needed load}, the total communication load in this example is given by $L_A=L_1+L_{[2:4]}=\frac{4171}{7260}$.

\subsection{General Scheme}\label{sec achievable load heterogeneous}
Consider a general $K$-node MapReduce computing system with computation load $\mathbf{m}=[m_1,\ldots,m_K]$ and function assignment $\mathbf{w}=[w_1,\ldots,w_K]$.

\subsubsection{Map phase design}
Similar to Section \ref{sec example}, the file allocation strategy has two steps. First, the $N$ input files are partitioned into $K$ disjoint batches, each computed by a distinct node; then, nodes with redundant computation capabilities compute the Map functions of more files from their non-computed files.

The main idea in the first step is to allocate the input files among all the nodes exclusively as equal as possible.  More specifically, define
\begin{align}
  l_k\triangleq\min\left\{m_k,a_k\right\},\textrm{ for }k\in[K]\label{eqn general scheme l definition}
\end{align}
with
\begin{align}
a_k\triangleq\left\{
\begin{array}{ll}
\frac{1}{K},&\textrm{ if }k=1,\\
\frac{1-\sum_{i=1}^{k-1}l_i}{K-k+1},&\textrm{ if }k\in[2:K].
\end{array}
\right.\label{eqn general scheme l definition 1}
\end{align}
It is easy to prove that $\sum_{k=1}^Kl_k=1$. Then, in the first step, we partition the $N$ input files into $K$ disjoint batches $\{\mathcal{N}_k:k\in[K]\}$. The $k$-th batch $\mathcal{N}_k$ has $l_kN$ files, and is allocated to node $k$.

The determination of $\{l_k:k\in[K]\}$ can be explained as follows. Consider an arbitrary node $k$. Given that node $i$, for $i\in[k-1]$, is allocated $l_iN$ files, there remain $(1-\sum_{i=1}^{k-1}l_i)N$ files need to be allocated to nodes $[k:K]$. If $m_k>a_k$, then we equally  partition these files into $K-k+1$ disjoint batches, and allocate them to nodes $[k:K]$; otherwise, we let node $k$ exhaust its computation capability to compute $m_kN$ files. Therefore, the number of files allocated to node $k$ is given by \eqref{eqn general scheme l definition}.

It is easy to prove that if $m_k\le a_k$ in \eqref{eqn general scheme l definition}, then $m_{k-1}\le a_{k-1}$, for $k\in[2:K]$. Define
\begin{align}
  r\triangleq \max\limits_{m_k\le a_k}k=\max\limits_{(K-k+1)m_k+\sum_{i=1}^{k-1}l_i\le1} k\label{eqn general define r}
\end{align}
as the largest index of nodes such that $m_k\le a_k$. We have $m_k\le a_k$ for nodes $[r]$, and $m_k>a_k$ for nodes $[r+1:K]$. Define $\xi\triangleq\sum_{k=1}^r m_k$, then we have $l_k=m_k$ for $k\in[r]$, and $l_k=\frac{1-\xi}{K-r}$ for $k\in[r+1:K]$.

After the first step, each node $k$, for $k\in[K]$, is still able to compute the Map functions of $(m_k-l_k)N$ more files from the remaining $(1-l_k)N$ files $\{f_1,\ldots,f_N\}\setminus\mathcal{N}_k$. Define $P_k\triangleq\frac{m_k-l_k}{1-l_k}$ as the surplus computation ratio of node $k$. In the second step, similar to Section \ref{sec example}, we further partition each batch $\mathcal{N}_k$, for $k\in[K]$, into $2^{K-1}$ sub-batches $\{\mathcal{N}_k^{\Psi}:\Psi\subseteq[K]\setminus\{k\}\}$. Each sub-batch $\mathcal{N}_k^{\Psi}$ has $l_k^{\Psi}N$ files with
\begin{align}
  l_k^{\Psi}\triangleq l_k\prod_{i\in\Psi}P_i\prod_{i\in[K]\setminus\{\Psi, k\}}(1-P_i),\label{eqn general file size}
\end{align}
and is further allocated to nodes in $\Psi\subseteq[K]\setminus\{k\}$. Note that $\mathcal{N}_k^{\Psi}=\emptyset$ and $l_k^{\Psi}=0$ if $\Psi\cap[r]\neq \emptyset$, since $P_i=0$ for $i\in[r]$. Here, the file allocation strategy in the second step is inspired by the decentralized cache placement \cite{decentralized,Wangsinong}.

By using the proposed two-step file allocation strategy, each node $k$ is allocated $l_kN$ files in the first step, and $l_i N\cdot P_k$ files from $\mathcal{N}_i$ for $i\in[K]\setminus\{k\}$ in the second step. It is easy to verify that each node $k$ computes the Map functions of $m_kN$ files in the Map phase, and satisfies its computation load. For each node $k$, we refer to the files selected in the first step, i.e., $\mathcal{N}_k$, as its \emph{compulsory} files, and refer to the files selected in the second step as its \emph{optional} files, denoted by $\mathcal{O}_k$. We have $\mathcal{M}_k=\mathcal{N}_k\cup\mathcal{O}_k$. Note that nodes $[r]$ have no optional file, i.e., $\mathcal{O}_k=\emptyset$ for $k\in[r]$, since $m_k=l_k$. Similar to Section  \ref{sec example}, we refer to nodes $[r]$ as LowCL nodes, and refer to nodes $[r+1:K]$ as HighCL nodes.


\subsubsection{Shuffle phase design}
In the Shuffle phase, each node $k$ needs the IVs of the other $(1-m_k)N$ files to compute the Reduce functions of its assigned functions $\mathcal{W}_k$, and should distribute the IVs computed from its compulsory files $\mathcal{N}_k$ to the requiring nodes. We first consider the communication to LowCL nodes $[r]$, and then consider the communication to HighCL nodes $[r+1:K]$.

\textbf{Communication to nodes $[r]$}: Consider an arbitrary node $i\in[r]$. Each node $k\in[K]\setminus\{i\}$ directly unicasts the IVs of the output functions $\mathcal{W}_i$ in its compulsory files $\mathcal{N}_k$ to node $i$, given by
\begin{align}
  \mathcal{V}_{k\rightarrow\{i\}}=\left\{v_{q,n}:\phi_q\in \mathcal{W}_i, f_n\in \mathcal{N}_k \right\}.\notag
\end{align}
The communication load from node $k$ to node $i$ is thus given by
\begin{align}
  L_{k\rightarrow\{i\}}=\frac{w_iQ\cdot l_kN\cdot T}{QNT}=w_il_k.\notag
\end{align}

After receiving its desired IVs computed from files $\{\mathcal{N}_k:k\in[K]\setminus\{i\}\}$, and combining the IVs computed locally from files $\mathcal{N}_i$, node $i$ can successfully compute the Reduce functions of its assigned functions $\mathcal{W}_i$ and obtain the output values.

Combining the communication from each node $k\in[K]\setminus\{i\}$ to each node $i\in[r]$, the sum communication load to nodes $[r]$ is given by
\begin{align}
  L_{[r]}=\sum_{i\in[r]}\sum_{k\in[K]\setminus\{i\}}w_il_k=\sum_{i\in[r]}w_i(1-l_i)=\sum_{i\in[r]}w_i(1-m_i).\label{eqn general node r needed load}
\end{align}

\textbf{Communication to nodes $[r+1:K]$}: Since each node $i\in[r+1:K]$ has the IVs computed from its optional files $\mathcal{O}_i$, each node $k\in[K]\setminus\{i\}$ only needs to send the IVs of the functions $\mathcal{W}_i$ in the rest of its compulsory files $\mathcal{N}_k$ to node $i$, given by
\begin{align}
  \left\{v_{q,n}:\phi_q\in \mathcal{W}_i, f_n\in \mathcal{N}_k\setminus\mathcal{O}_i \right\}.\notag
\end{align}
Note that the Map function of each file in $\mathcal{N}_k\setminus\mathcal{O}_i$ is possibly computed by nodes in $[r+1:K]\setminus\{i,k\}$ given the
second step of file allocation in the Map phase. Thus, similar to Section \ref{sec example}, coded multicasting opportunities can be exploited.

Consider an arbitrary node subset $\Psi\subseteq[r+1:K]$ and an arbitrary node $k\in[K]\setminus\Psi$. For each node $i\in\Psi$, the IVs
\begin{align}
\mathcal{V}_{k\rightarrow i}^{\Psi\setminus\{i\}}\triangleq\left\{v_{q,n}:\phi_q\in \mathcal{W}_i, f_n\in \mathcal{N}_k^{\Psi\setminus\{i\}}\right\}\notag
\end{align}
are needed by node $i$ and available at nodes $\Psi\setminus\{i\}$. The number of IVs in $\mathcal{V}_{k\rightarrow i}^{\Psi\setminus\{i\}}$ is given by
\begin{align}
  |\mathcal{V}_{k\rightarrow i}^{\Psi\setminus\{i\}}|=&w_iQl_kN\prod_{j\in\Psi\setminus\{i\}}P_j\cdot\prod_{j\in[r+1:K]\setminus\Psi,j\neq k}(1-P_j)\cdot(1-P_i)\notag\\
  =&l_kN\frac{w_iQ(1-P_i)}{P_i}\prod_{j\in\Psi}P_j\cdot\prod_{j\in[r+1:K]\setminus\Psi,j\neq k}(1-P_j).\notag
\end{align}
Using zero-padding and bit-wise XOR combining, node $k$ can directly multicast
\begin{align}
  \mathcal{V}_{k\rightarrow\Psi}=\bigoplus_{i\in\Psi}\mathcal{V}_{k\rightarrow i}^{\Psi\setminus\{i\}}\notag
\end{align}
to nodes $\Psi$, and each node $i\in\Psi$ can successfully obtain its desired IVs $\mathcal{V}_{k\rightarrow i}^{\Psi\setminus\{i\}}$ since it already has IVs $\{\mathcal{V}_{k\rightarrow j}^{\Psi\setminus\{j\}}:j\in\Psi\setminus\{i\}\}$. The communication load is determined by the largest number of needed IVs among nodes in $\Psi$, given by
\begin{align}
L_{k\rightarrow\Psi}=\max_{i\in\Psi}\frac{|\mathcal{V}_{k\rightarrow i}^{\Psi\setminus\{i\}}|}{QNT}=l_k\prod_{j\in\Psi}P_j\cdot\!\!\prod_{j\in[r+1:K]\setminus\Psi,j\neq k}(1-P_j)\!\cdot\! \max_{i\in\Psi}\frac{w_i(1-P_i)}{P_i}.\notag
\end{align}

Based on this shuffle strategy, each node $i\!\in\![r+1\!:\!K]$ can obtain its needed IVs $\mathcal{V}_{k\rightarrow i}^{\Psi}$ from coded message $\mathcal{V}_{k\rightarrow\Psi\cup\{i\}}$, for $k\in[K]\setminus\{i\},\Psi\!\subseteq\![r+1\!:\!K]\!\setminus\!\{i,\!k\}$. Combining the IVs computed locally from files $\mathcal{M}_i$, node $i$ can successfully compute the Reduce functions of its assigned functions $\mathcal{W}_i$ and obtain the output values.

Now let us calculate the required communication load for the proposed shuffle strategy. Reorder nodes $[r+1:K]$ in descending order of the value $\frac{w_k(1-P_k)}{P_k}$ such that the $i$-th node $s_i$ has the $i$-th largest value of $\frac{w_k(1-P_k)}{P_k}$, i.e.,
\begin{align}
  \frac{w_{s_1}(1-P_{s_1})}{P_{s_1}}\ge\frac{w_{s_2}(1-P_{s_2})}{P_{s_2}}\ge\ldots\ge\frac{w_{s_{K-r}}(1-P_{s_{K-r}})}{P_{s_{K-r}}}.\notag
\end{align}
Define node set $\mathcal{S}_{[a:b]}\triangleq[s_a,s_{a+1},\ldots,s_{b-1},s_{b}]$, for $a\le b, a,b\in\mathbb{Z}$. Then, for an arbitrary node $k\in[r]$, the total communication load sent from node $k$ to nodes $[r+1:K]$ is given by
\begin{align}
  L_{[r+1:K]}^{k}=&\sum_{\Psi\subseteq[r+1:K]}L_{k\rightarrow\Psi}\notag\\
  =&\sum_{i=1}^{K-r}\sum_{\substack{\Psi\subseteq[r+1:K],s_i\in\Psi\\s_j\notin\Psi,\forall j\in[i-1]}}L_{k\rightarrow\Psi}\notag\\
  =&\sum_{i=1}^{K-r}\sum_{\substack{\Psi\subseteq[r+1:K],s_i\in\Psi\\s_j\notin\Psi,\forall j\in[i-1]}}l_k\prod_{j\in\Psi}P_j\cdot\!\!\prod_{j\in[r+1:K]\setminus\Psi}(1-P_j)\!\cdot\! \frac{w_{s_i}(1-P_{s_i})}{P_{s_i}}\notag\\
  =&\sum_{i=1}^{K-r}l_k\frac{w_{s_i}(1-P_{s_i})}{P_{s_i}}P_{s_i}\prod_{j\in[i-1]}(1-P_{s_j})\cdot\sum_{\Psi\subseteq \mathcal{S}_{[i+1,K-r]}}\prod_{j\in\Psi}P_j\cdot\!\!\prod_{j\in \mathcal{S}_{[i+1,K-r]}\setminus\Psi}(1-P_j)\notag\\
  =&\sum_{i=1}^{K-r}l_kw_{s_i}\prod_{j\in[i]}(1-P_{s_j})\label{eqn general node K-r load from r}
\end{align}
For an arbitrary node $s_k\in[r+1:K]$, the total communication load sent from node $s_k$ to nodes $[r+1:K]\setminus\{s_k\}$ is given by
\begin{align}
  &L_{[r+1:K]\setminus\{s_k\}}^{s_k}\notag\\
  =&\sum_{\Psi\subseteq[r+1:K]\setminus\{s_k\}}L_{s_k\rightarrow\Psi}\notag\\
  =&\sum_{i=1}^{k-1}\sum_{\substack{\Psi\subseteq[r+1:K]\setminus\{s_k\},s_i\in\Psi\\s_j\notin\Psi,\forall j\in[i-1]}}L_{s_k\rightarrow\Psi}+\sum_{i=k+1}^{K-r}\sum_{\substack{\Psi\subseteq[r+1:K]\setminus\{s_k\},s_i\in\Psi\\s_j\notin\Psi,\forall j\in[i-1]}}L_{s_k\rightarrow\Psi}\notag\\
  =&\sum_{i=1}^{k-1}\sum_{\substack{\Psi\subseteq[r+1:K]\setminus\{s_k\},s_i\in\Psi\\s_j\notin\Psi,\forall j\in[i-1]}}l_{s_k}\prod_{j\in\Psi}P_j\cdot\!\!\prod_{j\in[r+1:K]\setminus\Psi,j\neq s_k}(1-P_j)\!\cdot\! \frac{w_{s_i}(1-P_{s_i})}{P_{s_i}}\notag\\
  &+\sum_{i=k+1}^{K-r}\sum_{\substack{\Psi\subseteq[r+1:K]\setminus\{s_k\},s_i\in\Psi\\s_j\notin\Psi,\forall j\in[i-1]}}l_{s_k}\prod_{j\in\Psi}P_j\cdot\!\!\prod_{j\in[r+1:K]\setminus\Psi,j\neq s_k}(1-P_j)\!\cdot\! \frac{w_{s_i}(1-P_{s_i})}{P_{s_i}}\notag\\
  =&\sum_{i=1}^{k-1}l_{s_k}\frac{w_{s_i}(1-P_{s_i})}{P_{s_i}}P_{s_i}\prod_{j\in[i-1]}(1-P_{s_j})\cdot\sum_{\Psi\subseteq \mathcal{S}_{[i+1,K-r]}\setminus\{s_k\}}\prod_{j\in\Psi}P_j\cdot\!\!\prod_{j\in \mathcal{S}_{[i+1,K-r]}\setminus\Psi,j\neq s_k}(1-P_j)\notag\\
  &+\sum_{i=k+1}^{K-r}l_{s_k}\frac{w_{s_i}(1-P_{s_i})}{P_{s_i}}P_{s_i}\prod_{j\in[i-1],j\neq k}(1-P_{s_j})\cdot\sum_{\Psi\subseteq \mathcal{S}_{[i+1,K-r]}}\prod_{j\in\Psi}P_j\cdot\!\!\prod_{j\in \mathcal{S}_{[i+1,K-r]}\setminus\Psi}(1-P_j)\notag\\
  =&\sum_{i=1}^{k-1}l_{s_k}w_{s_i}\prod_{j\in[i]}(1-P_{s_j})+\sum_{i=k+1}^{K-r}l_{s_k}w_{s_i}\frac{1}{1-P_{s_k}}\prod_{j\in[i]}(1-P_{s_j})\label{eqn general node K-r load from K-r}
\end{align}
Here, the derivations in \eqref{eqn general node K-r load from r} and \eqref{eqn general node K-r load from K-r} are similar to that in \cite[Theorem 3]{Wangsinong}. Summing up \eqref{eqn general node K-r load from r} for all $k\in[r]$ and \eqref{eqn general node K-r load from K-r} for all $s_k\in[r+1:K]$, the total communication load to nodes $[r+1:K]$ is given by
\begin{align}
  L_{[r+1:K]}=&\sum_{k\in[r]}L_{[r+1:K]}^{k}+\sum_{k\in[K-r]}L_{[r+1:K]\setminus\{s_k\}}^{s_k}\notag\\
  =&\sum_{k\in[r]}\sum_{i=1}^{K-r}l_kw_{s_i}\prod_{j\in[i]}(1-P_{s_j})+\sum_{k\in[K-r]}\sum_{i=1}^{k-1}l_{s_k}w_{s_i}\prod_{j\in[i]}(1-P_{s_j})\notag\\
  &+\sum_{k\in[K-r]}\sum_{i=k+1}^{K-r}l_{s_k}w_{s_i}\frac{1}{1-P_{s_k}}\prod_{j\in[i]}(1-P_{s_j})\notag\\
  =&\sum_{i\in[K-r]}w_{s_i}\prod_{j\in[i]}(1-P_{s_j})\cdot\left[\sum_{k\in[r]}l_k+\sum_{k\in[i-1]}\frac{l_{s_k}}{1-P_{s_k}}+\sum_{k\in[i+1:K-r]}l_{s_k}\right]\notag\\
  =&\sum_{i\in[K-r]}w_{s_i}\prod_{j\in[i]}(1-P_{s_j})\cdot\left[\xi+\frac{1-\xi}{K-r}\sum_{k\in[i-1]}\frac{1}{1-P_{s_k}}+(K-r-i)\frac{1-\xi}{K-r}\right],\label{eqn general node K-r load}
\end{align}
where the last equality comes from the fact that $l_k=m_k$ for $k\in[r]$ and $l_k=\frac{1-\xi}{K-r}$ for $k\in[r+1:K]$. Combining \eqref{eqn general node r needed load} and \eqref{eqn general node K-r load}, the total communication load in the Shuffle phase is given by
\begin{align}
  L=&\sum_{k\in[r]}w_k(1-m_k)\notag\\
  &+\sum_{i\in[K-r]}w_{s_i}\prod_{j\in[i]}(1-P_{s_j})\cdot\left[\xi+\frac{1-\xi}{K-r}\sum_{k\in[i-1]}\frac{1}{1-P_{s_k}}+(K-r-i)\frac{1-\xi}{K-r}\right].\notag
\end{align}

\subsection{Achievable Communication Load}
The achievable communication load in a general heterogeneous MapReduce computing system is formally stated in the following theorem.

\vspace{-5pt}
\begin{theorem}\label{thm achievable load arbitrary}
For a heterogeneous MapReduce computing system with $K$ nodes, computation load $\mathbf{m}=[m_1,\ldots,m_K]$, and function assignment $\mathbf{w}=[w_1,\ldots,w_K]$, an achievable communication load is given by
\begin{align}
  L_A(\mathbf{m},\!\mathbf{w})\!\triangleq\!\sum_{k=1}^r w_k(1\!-\!m_k)+\!\sum_{k=1}^{K-r}w_{s_k}\prod_{i=1}^k(1\!-\!P_{s_i})\cdot\left[\xi+(K\!-\!r\!-\!k)\frac{1\!-\!\xi}{K\!-\!r}\!+\!\frac{1\!-\!\xi}{K\!-\!r}\sum_{i=1}^{k-1}\frac{1}{1\!-\!P_{s_i}}\right]\label{eqn achievable load arbitrary}
\end{align}
where
\begin{align}
  r\triangleq \max\limits_{(K-k+1)m_k+\sum_{i=1}^{k-1}l_i\le1} k\label{eqn r define}
\end{align}
with $l_1=\min\{m_1,\frac{1}{K}\}$, $l_k=\min\{m_k,\frac{1-\sum_{i=1}^{k-1}l_i}{K-k+1}\}$ for $k\in[2:K]$; $\xi\triangleq\sum_{k=1}^{r}m_k$; $P_{k}\triangleq\frac{m_k-l_k}{1-l_k}$ for $k\in[K]$; and $\{s_1,\ldots,s_{K-r}\}$ is the re-ordered indices of nodes $[r+1:K]$ in descending order of the value $\frac{w_k(1-P_k)}{P_k}$, i.e.,
\begin{align}
  \frac{w_{s_1}(1-P_{s_1})}{P_{s_1}}\ge\frac{w_{s_2}(1-P_{s_2})}{P_{s_2}}\ge\ldots\ge\frac{w_{s_{K-r}}(1-P_{s_{K-r}})}{P_{s_{K-r}}}.\notag
\end{align}
\end{theorem}
\vspace{-5pt}

\begin{remark}[Homogeneous system]
When each node has the same computation load $m_k=m$ and the same function assignment $w_k=\frac{1}{K}$, the communication load in Theorem \ref{thm achievable load arbitrary} reduces to

\noindent 1) if $m=\frac{1}{K}$:
\begin{align}
L_{A-1}\!=\!\sum_{k=1}^K \frac{1}{K}(1\!-\!\frac{1}{K})\!=\!\frac{K\!-\!1}{K},\label{eqn achievable load homogeneous 1}
\end{align}

\noindent 2) if $m>\frac{1}{K}$:
\begin{align}
L_{A-2}=&\sum_{k=1}^{K}\frac{1}{K}\left(\frac{1-m}{1-\frac{1}{K}}\right)^k\left[\frac{K-k}{K}+\frac{k-1}{K}\frac{1-\frac{1}{K}}{1-m}\right]\notag\\
=&\frac{1}{K}\frac{1-m}{1-\frac{1}{K}}\frac{1-\left(\frac{1-m}{1-\frac{1}{K}}\right)^{K-1}}{1-\frac{1-m}{1-\frac{1}{K}}}\notag\\
\le&\frac{1-m}{Km-1}.\label{eqn achievable load homogeneous 2}
\end{align}
Compared to the optimal load $L^*_{\textrm{Hom}}=\frac{1-m}{Km}$, for $Km\in[K]$, obtained in \cite{LiSongZeMapReduce} for the homogeneous computing system,  our achievable load is the same as theirs when $m=\frac{1}{K}$, and close to theirs when $m>\frac{1}{K}$. In specific, when $Km\in[2:K]$, the multiplicative gap between our achievable load \eqref{eqn achievable load homogeneous 2} and the optimal load $L^*_{\textrm{Hom}}$ is upper bounded by $\frac{L_A}{L^*_{\textrm{Hom}}}\le\frac{Km}{Km-1}\le2$.
\end{remark}
\vspace{-5pt}



\section{Function Assignments}
While most works consider even function assignment $\mathbf{w}_{\textrm{Even}}\!=\![\frac{1}{K}]^{1\times K}$, the number of output functions assigned to each node is generally related to the number of files it is allocated. More specifically, when a node is allocated more input files, indicating that it has better storage and computation capabilities, it is also assigned more output functions so as to reduce the communication load in the Shuffle phase as well as the overall computation latency. This provides us opportunities to further reduce the communication load.

In this section, we propose two function assignments, i.e., the \emph{computation-aware} and the \emph{shuffle-aware} function assignments. Then, we will compare our achievable communication loads by using these two function assignments with the results in \cite{LiSongZeMapReduce,JimingyueCascade,JimingyueNoncascade,Wangchengwei}, and  present some discussions.


\subsection{Computation-aware function assignment}\label{sec computation aware}
The computation-aware function assignment aims to balance the function assignment among nodes according to their computation capabilities so as to reduce the overall computation latency. Since the computation capability of each node can be reflected by its computation load in the Map phase, a natural way is to let the number of output functions assigned to each node be proportional to its computation load. That is, each node $k$ computes $w_kQ$ output functions, where $w_k=\frac{m_k}{\sum_{i\in[K]}\! m_i}$. We refer to
\begin{align}
\mathbf{w}_{\textrm{Com}}(\mathbf{m})\!\triangleq\!\Big[\frac{m_1}{\sum_{k\in[K]}m_k},\ldots,\frac{m_K}{\sum_{k\in[K]}m_k}\Big]\label{eqn computation aware assign}
\end{align}
as the computation-aware function assignment. Substituting \eqref{eqn computation aware assign} into \eqref{eqn achievable load arbitrary}, the achievable communication load is given in the following theorem.

\vspace{-5pt}
\begin{theorem}[Computation-aware function assignment]\label{thm achievable computation aware}
For a heterogeneous MapReduce computing system with $K$ nodes, computation load $\mathbf{m}=[m_1,\ldots,m_K]$, and computation-aware function assignment $\mathbf{w}_{\textrm{Com}}(\mathbf{m})$, an achievable communication load is given by
\begin{align}
  L_A(\mathbf{m},\mathbf{w}_{\textrm{Com}}(\mathbf{m}))\triangleq&\sum_{k=1}^{r}\frac{m_k}{\sum_{i\in[K]}m_i}(1-m_k)\!+\!\!\sum_{k=r+1}^{K}\!\frac{m_k}{\sum_{i\in[K]}m_i}\!\prod_{i=r+1}^{k}\!(1-P_i)\cdot\notag\\
  &\qquad\qquad\left[\xi+(K-k)\frac{1-\xi}{K-r}+\frac{1-\xi}{K-r}\sum_{i=r+1}^{k-1}\frac{1}{1-P_i}\right]\label{eqn achievable load computation aware}
\end{align}
with $r,\xi,\{P_k:k\in[K]\}$ defined in Theorem \ref{thm achievable load arbitrary}.
\end{theorem}
\vspace{-6pt}



\subsection{Shuffle-aware function assignment}\label{sec shuffle aware}
The shuffle-aware function assignment aims to reduce the traffic load in the Shuffle phase when $\sum_{k\in[K]}m_k>1$ by properly assigning output functions to HighCL nodes $[r+1:K]$. Note that when $\sum_{k\in[K]}m_k>1$, we have $r+1\le K$. We first use the example in Section \ref{sec example} to illustrate the assignment to HighCL nodes.  It can be seen in \eqref{eqn example one set load} that the communication load to an arbitrary node set $\Psi\subseteq[2:4]$ is determined by the largest number of  needed IVs among nodes in $\Psi$ due to zero-padding. To minimize the performance loss caused by zero-padding, we let $\frac{w_k(1-P_k)}{P_k}$ be equal for each node $k\in\Psi$. Traversing all $\Psi\subseteq[2:4]$, we have $\frac{w_2(1-P_2)}{P_2}\!=\!\frac{w_3(1-P_3)}{P_3}\!=\!\frac{w_4(1-P_4)}{P_4}$. In general, to avoid zero-padding in the communication to HighCL nodes $[r+1:K]$, the function assignment should satisfy $\frac{w_{r+1}(1-P_{r+1})}{P_{r+1}}=\cdots=\frac{w_K(1-P_K)}{P_K}$.

Now, let us consider the function assignment to LowCL nodes $[r]$. Note that the number of IVs computed by each LowCL node in $[r]$ is less than that computed by each HighCL node in $[r+1:K]$. Moreover, the proposed shuffle strategy adopts unicasting, instead of coded multicasting, to deliver the required IVs to LowCL nodes $[r]$. Therefore, to avoid the communication to these LowCL nodes becomes the bottleneck of the shuffle phase, we simply do not assign any output function to the LowCL nodes $[r]$, i.e., $w_k=0,\forall k\in[r]$.

By using the above strategy, the function assignment can be easily computed, and is given by
\begin{align}
  \mathbf{w}_{\textrm{Shu}}(\mathbf{m})\!\triangleq\!\left[\underbrace{0,\ldots,0}_{r \textrm{ zeros}},\frac{\frac{P_{r+1}}{1-P_{r+1}}}{\sum_{k=r+1}^K\!\frac{P_k}{1-P_k}},\ldots,\frac{\frac{P_{K}}{1-P_{K}}}{\sum_{k=r+1}^K\!\frac{P_k}{1-P_k}}\right]\!,\label{eqn shuffle aware assignment}
\end{align}
which is referred to as the shuffle-aware function assignment. Similar to the computation-aware function assignment \eqref{eqn computation aware assign} and the function assignment in \cite{JimingyueNoncascade}, it can be seen in \eqref{eqn shuffle aware assignment} that, among nodes $[r+1:K]$, those with higher computation load are assigned more output functions. However, unlike \eqref{eqn computation aware assign} and \cite{JimingyueNoncascade}, nodes $[r]$ do not compute output functions in this function assignment. Substituting \eqref{eqn shuffle aware assignment} into \eqref{eqn achievable load arbitrary}, the achievable communication load is given in the following theorem.

\vspace{-5pt}
\begin{theorem}[Shuffle-aware function assignment]\label{thm achievable shuffle aware}
For a heterogeneous MapReduce computing system with $K$ nodes, computation load $\mathbf{m}=[m_1,\ldots,m_K]$ with $\sum_{k\in[K]}m_k>1$, and shuffle-aware function assignment $\mathbf{w}_{\textrm{Shu}}(\mathbf{m})$, an achievable communication load is given by
\begin{align}
  L_{A}(\mathbf{m},\mathbf{w}_{\textrm{Shu}}(\mathbf{m}))\triangleq&\frac{1}{\sum_{k=r+1}^{K}\frac{P_k}{1-P_k}}\left[1-\xi\prod_{k=r+1}^K(1-P_k)-\frac{1\!-\!\xi}{K\!-\!r}\prod_{k=r+1}^K(1\!-\!P_k)\sum_{k=r+1}^K\frac{1}{1\!-\!P_k}\right]\label{eqn achievable load shuffle aware}
\end{align}
with $r,\xi,\{P_k:k\in[K]\}$ defined in Theorem \ref{thm achievable load arbitrary}.
\end{theorem}
\vspace{-6pt}

\subsection{Comparison with other works}
Given a heterogeneous MapReduce computing system with computation load $\mathbf{m}$, we can define an equivalent homogeneous system where the computation load at each node equals the average computation load $\bar{m}=\frac{\sum_{k}m_k}{K}$ in the heterogeneous system, and the function assignment is even with $\mathbf{w}_{\textrm{Even}}=[\frac{1}{K}]^{1\times K}$. According to \cite{LiSongZeMapReduce}, the optimal communication load $L^*_{\textrm{Hom}}(\bar{m})$ in this homogeneous system is given by the lower convex envelope of points $(\bar{m},\frac{1-\bar{m}}{K\bar{m}})$ for $\bar{m}\in[\frac{1}{K},\frac{2}{K},\ldots,1]$. The next corollary shows the multiplicative gap between our achievable communication load in the heterogeneous system and $L^*_{\textrm{Hom}}(\bar{m})$ in the equivalent homogeneous system.


\vspace{-5pt}
\begin{corollary}\label{coro gap aware function assignment}
For a heterogeneous MapReduce computing system with $K$ nodes and computation load $\mathbf{m}$, when $\bar{m}<0.55$, the multiplicative gap between our achievable communication load $L_A(\mathbf{m},\mathbf{w}_{\textrm{Com}}(\mathbf{m}))$ using computation-aware function assignment and the optimal load $L^*_{\textrm{Hom}}(\bar{m})$ in the equivalent homogeneous system is within 115; when $\bar{m}\ge0.55$, the multiplicative gap between our achievable communication load $L_A(\mathbf{m},\mathbf{w}_{\textrm{Shu}}(\mathbf{m}))$ using shuffle-aware function assignment and the optimal load $L^*_{\textrm{Hom}}(\bar{m})$ in the equivalent homogeneous system is within 115.
\end{corollary}
\vspace{-5pt}

The proof of Corollary \ref{coro gap aware function assignment} is in Appendix A. Corollary \ref{coro gap aware function assignment} implies that, even with heterogeneous computation load, by designing proper function assignment, the achievable communication load is still within a constant multiplicative gap to the optimum in the equivalent homogeneous system.

\begin{figure}[!tbp]
\centering
\includegraphics[scale=0.4]{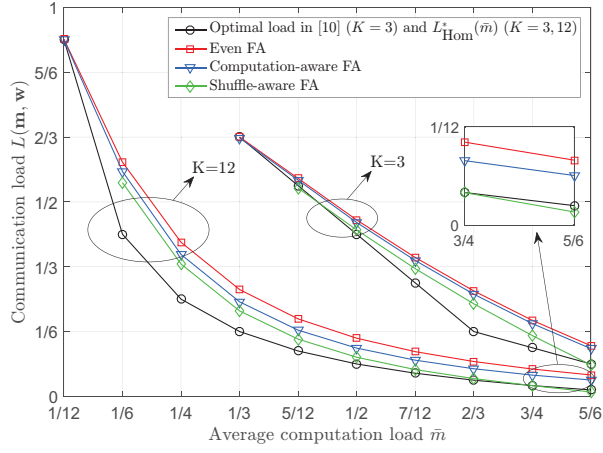}
\vspace{-5pt}
\caption{Communication load $L$ with $K=3$ and $K=12$.}\label{Fig numerical comparison}
\vspace{-10pt}
\end{figure}

Fig. \ref{Fig numerical comparison} plots the achievable communication loads of \cite{Wangchengwei} and our work with respect to the average computation load $\bar{m}$ in the heterogeneous MapReduce computing systems with $K=3$ and $K=12$ as well as the optimal communication load $L^*_{\textrm{Hom}}(\bar{m})$ in the equivalent homogeneous systems. The computation load in \cite{Wangchengwei} and our work is $\mathbf{m}\!=\!\bar{m}\cdot[0.9,1,1.1]$ for $K\!=\!3$ and our work also considers $\mathbf{m}\!=\!\bar{m}\cdot[0.7,0.8,0.9,0.9,0.9,1,1,1.05,1.1,1.1,1.15,1.15]$ for $K\!=\!12$. Note that \cite{Wangchengwei} obtains the optimal communication load when $K=3$ for even function assignment $\mathbf{w}_{\textrm{Even}}\!=\![\frac{1}{K}]^{1\times K}$, while we plot our results for even function assignment, computation-aware function assignment $\mathbf{w}_{\textrm{Com}}(\mathbf{m})$, and shuffle-aware function assignment $\mathbf{w}_{\textrm{Shu}}(\mathbf{m})$. Among our achievable results, the shuffle-aware function assignment achieves the smallest communication load, while the even function assignment achieves the largest. All these loads are close to the optimal load $L_{\textrm{Hom}}^*$ in the equivalent homogeneous system and the optimal load in \cite{Wangchengwei}. When $\bar{m}>0.75$ and $K=12$, our shuffle-aware function assignment achieves smaller communication load than $L_{\textrm{Hom}}^*$, because: 1) coded multicasting opportunities are sufficiently exploited by this function assignment; 2) nodes with higher computation load are assigned more output functions and less communication is needed to satisfy the requests of these nodes.

Table \ref{table comparison} shows the achievable communication loads of \cite{JimingyueCascade,JimingyueNoncascade} and our results with four function assignments for certain $\mathbf{m}$ in the MapReduce systems with $K=12$. The communication load in \cite{JimingyueCascade} is the largest because each output function is computed by multiple nodes. Note that the heterogeneous function assignment in \cite{JimingyueNoncascade} is tailored for its coded multicasting strategy in the Shuffle phase, which also avoids zero-padding in the generation of coded messages, similar to our shuffle-aware function assignment. Compared to \cite{JimingyueNoncascade}, our scheme by using the function assignment in \cite{JimingyueNoncascade} achieves smaller communication load for $\mathbf{m}_1$, and our shuffle-aware function assignment achieves smaller communication loads for both $\mathbf{m}$, because our shuffle strategy exploits coded multicasting opportunities for each subset of HighCL nodes (containing at least two nodes) in the Shuffle phase while  \cite{JimingyueNoncascade} only exploits them for some subsets of nodes.


\begin{table}[]
\centering
\renewcommand\tabcolsep{4pt}
\renewcommand{\arraystretch}{0.95}
\caption{Communication load $L$ with $K=12$: 1) $\mathbf{m}_1$: $m_k\!=\!\frac{1}{6}$ for $k\!\in\![1\!:\!6]$, $m_k\!=\!\frac{1}{3}$ for $k\!\in\![7\!:\!12]$; 2) $\mathbf{m}_2$: $m_k\!=\!\frac{1}{6}$ for $k\!\in\![1\!:\!6]$, $m_k\!=\!\frac{1}{2}$ for $k\!\in\![7\!:\!12]$.}\label{table comparison}
\vspace{-5pt}
\begin{tabular}{|c|c|c|}
\hline
\backslashbox{Scheme}{$\quad L\qquad$}{$\quad\mathbf{m}\qquad$}                                   & $\mathbf{m}_1$ & $\mathbf{m}_2$ \\ \hline
\cite{JimingyueCascade}                                                                             & 0.528                                                                                                                                  & 0.497                                                                                                                                  \\ \hline
\cite{JimingyueNoncascade}                                                                             & 0.357                                                                                                                                  & 0.185                                                                                                                                  \\ \hline
Even FA            & 0.448                                                                                                                                  & 0.397                                                                                                                                  \\ \hline
Computation-aware FA & 0.371                                                                                                                                  & 0.255                                                                                                                                  \\ \hline
Shuffle-aware FA     & 0.315                                                                                                                                  & 0.175                                                                                                                                  \\ \hline
FA in \cite{JimingyueNoncascade}    & 0.349                                                                                                                                  & 0.208                                                                                                                                  \\ \hline
\end{tabular}
\vspace{-20pt}
\end{table}

\subsection{Discussion on the required numbers of input files and output functions}
In our proposed scheme, we need to ensure that the number of input files in each sub-batch and the number of output functions assigned to each node are integers. Though it is very challenging to derive the exact numbers of input files and output functions required for our scheme in the general heterogeneous system, we can still provide some analysis on the magnitude of the required numbers.

In our file allocation strategy, if $r>0$, we can prove that $l_1\le \cdots \le l_K$, with $l_k$ defined in Theorem 1. Then, the least number of input files in the sub-batches is given by $l_1N\prod_{k\in[r+1:K]}\min\{P_k,1-P_k\}$, where $P_k=\frac{m_k-l_k}{1-l_k}$ for $k\in[K]$. To ensure this number to be an integer, i.e.,
\begin{align}
l_1N\prod_{k\in[r+1:K]}\min\{P_k,1-P_k\}=z\in\mathbb{Z}^+,\notag
\end{align}
$N$ should satisfy $N=\frac{z}{l_1\prod_{k\in[r+1:K]}\min\{P_k,1-P_k\}}$. Therefore, the input file number $N$ should scale with the multiple of $\frac{1}{l_1\prod_{k\in[r+1:K]}\min\{P_k,1-P_k\}}$. If $r=0$, which implies $l_1=\cdots=l_K=\frac{1}{K}$, the least number of input files in the sub-batches is given by $\frac{1}{K}N\frac{\prod_{k\in[K]}\min\{P_k,1-P_k\}}{\max_{k\in[K]}\min\{P_k,1-P_k\}}$. Then, similar to the case when $r>0$, the input file number $N$ should scale with the multiple of $\frac{K\max_{k\in[K]}\min\{P_k,1-P_k\}}{\prod_{k\in[K]}\min\{P_k,1-P_k\}}$.

In our computation-aware function assignment, the least number of output functions assigned to nodes is given by $\frac{m_1}{\sum_{k\in[K]}m_k}$. Therefore, similar to file allocation, $Q$ should scale with the multiple of $\frac{\sum_{k\in[K]}m_k}{m_1}$. In our shuffle-aware function assignment, the least number of output functions assigned to nodes $[r+1:K]$ is given by $\frac{\frac{P_{r+1}}{1-P_{r+1}}}{\sum_{k=r+1}^K\!\frac{P_k}{1-P_k}}$. Therefore, similar to file allocation, $Q$ should scale with the multiple of $\frac{\sum_{k=r+1}^K\!\frac{P_k}{1-P_k}}{\frac{P_{r+1}}{1-P_{r+1}}}$.

Table \ref{table file function number} lists the least numbers required for input files and output functions in \cite{Wangchengwei,LiSongZeMapReduce,JimingyueCascade,JimingyueNoncascade} and our scheme in the MapReduce system considered in Section IV-C. It can be seen that the numbers required for output functions in our function assignment strategies are relatively close to existing works, and our computation-aware function assignment requires less number of output functions than those in \cite{JimingyueCascade,JimingyueNoncascade}, but the number required for input files in our scheme is much larger than existing works. This is because our file allocation strategy is inspired by the decentralized cache placement, and the proportion of each sub-batch in the entire input files is given by a product of a sequence as in \eqref{eqn general file size} which becomes very small when $K$ is large. Thus, to guarantee the number of input files in each sub-batch to be an integer, our scheme requires a large number of input files $N$. However, note that our scheme is applicable to the MapReduce computing system for any given node number $K\ge2$, any given computation load $\mathbf{m}$, and any given function assignment $\mathbf{w}$ (with $\sum_{k}w_k=1$), which is more general than those considered in \cite{Wangchengwei,LiSongZeMapReduce,JimingyueCascade,JimingyueNoncascade}.

\begin{table}[]
\centering
\renewcommand\tabcolsep{1pt}
\caption{The least numbers required for input files and output functions}
\label{table file function number}
\begin{tabular}{|c|c|c|c|c|}
\hline
Node number $K$         & Computation load $\mathbf{m}$                                                                                                                         & Scheme                                                                                          & \begin{tabular}[c]{@{}c@{}}Least number required\\ for input files\end{tabular} & \begin{tabular}[c]{@{}c@{}}Least number required \\ for output functions\end{tabular} \\ \hline
\multirow{4}{*}{$K=3$}  & $\mathbf{m}=[\frac{2}{3},\frac{2}{3},\frac{2}{3}]$                                                                                                    & \cite{LiSongZeMapReduce}                                                                        & 3                                                                               & 3                                                                                     \\ \cline{2-5}
                        & \multirow{3}{*}{$\mathbf{m}=[\frac{3}{5},\frac{2}{3},\frac{11}{15}]$}                                                                                 & \cite{Wangchengwei}                                                                             & 15                                                                              & 3                                                                                     \\ \cline{3-5}
                        &                                                                                                                                                       & Computation-aware FA & 150                                                                             & 30                                                                                    \\ \cline{3-5}
                        &                                                                                                                                                       & Shuffle-aware FA     & 150                                                                             & 19                                                                                    \\ \hline
\multirow{5}{*}{$K=12$} & $\mathbf{m}\!=\![\frac{1}{4}]^{1\times12}$                                                                                                       & \cite{LiSongZeMapReduce}                                                                        & 220                                                                             & 12                                                                                    \\ \cline{2-5}
                        & \multirow{4}{*}{$\mathbf{m}_1$} & \cite{JimingyueCascade}                                                                         & 54                                                                              & 54                                                                                    \\ \cline{3-5}
                        &                                                                                                                                                       & \cite{JimingyueNoncascade}                                                                      & 54                                                                              & 42                                                                                    \\ \cline{3-5}
                        &                                                                                                                                                       & Computation-aware FA & $12\cdot 11^{11}$                                                               & 18                                                                                    \\ \cline{3-5}
                        &                                                                                                                                                       & Shuffle-aware FA     & $12\cdot 11^{11}$                                                               & 114                                                                                   \\ \hline
\multirow{5}{*}{$K=12$} & $\mathbf{m}\!=\![\frac{1}{3}]^{1\times 12}$                                                                                                      & \cite{LiSongZeMapReduce}                                                                        & 495                                                                             & 12                                                                                    \\ \cline{2-5}
                        & \multirow{4}{*}{$\mathbf{m}_2$} & \cite{JimingyueCascade}                                                                         & 48                                                                              & 48                                                                                    \\ \cline{3-5}
                        &                                                                                                                                                       & \cite{JimingyueNoncascade}                                                                      & 48                                                                              & 36                                                                                    \\ \cline{3-5}
                        &                                                                                                                                                       & Computation-aware FA & $12\cdot 11^{11}$                                                               & 24                                                                                    \\ \cline{3-5}
                        &                                                                                                                                                       & Shuffle-aware FA     & $12\cdot 11^{11}$                                                               & 168                                                                                   \\ \hline
\end{tabular}
\vspace{-20pt}
\end{table}

\setcounter{subsection}{0}
\section*{Appendix A: Proof of Corollary \ref{coro gap aware function assignment}}
The optimal communication load $L^*_{\textrm{Hom}}(\bar{m})$ in the equivalent homogeneous system is given by the lower convex envelope of points $(\bar{m},\frac{1-\bar{m}}{K\bar{m}})$ for $\bar{m}\in[\frac{1}{K},\frac{2}{K},\ldots,1]$. Since $\frac{1-\bar{m}}{K\bar{m}}$ is a convex and decreasing function of $\bar{m}$, we have $L^*_{\textrm{Hom}}(\bar{m})\ge \frac{1-\bar{m}}{K\bar{m}}$. In the following, we consider two cases to prove Corollary \ref{coro gap aware function assignment}: 1) $\bar{m}=\frac{\sum_{k\in[K]}m_k}{K}<0.55$ and the communication load is achieved by $L_A(\mathbf{m},\mathbf{w}_{\textrm{Com}}(\mathbf{m}))$; 2) $\bar{m}\ge0.55$ and the communication load is achieved by $L_A(\mathbf{m},\mathbf{w}_{\textrm{Shu}}(\mathbf{m}))$.

\subsection{$\bar{m}<0.55$}
We first present an information-theoretical lower bound of the minimum communication load $L^*(\mathbf{m},\mathbf{w})$ for arbitrary computation load $\mathbf{m}$ and arbitrary function assignment $\mathbf{w}$, whose proof is in Appendix B.

\begin{lemma}\label{thm converse load arbitrary}
For a heterogeneous MapReduce computing system with $K$ nodes, computation load $\mathbf{m}=[m_1,\ldots,m_K]$, and function assignment $\mathbf{w}=[w_1,\ldots,w_K]$, the minimum communication load $L^*(\mathbf{m},\mathbf{w})$ is lower bounded by
\begin{align}
L^*(\mathbf{m},\mathbf{w})\ge L_{Lower}(\mathbf{m},\mathbf{w})\triangleq\max_{\mathcal{T}\subseteq[K]}(1-\sum_{k\in\mathcal{T}}m_k)\sum_{k\in\mathcal{T}}w_k.\label{eqn converse load arbitrary}
\end{align}
\end{lemma}

Comparing Theorem \ref{thm achievable computation aware} and Lemma \ref{thm converse load arbitrary}, the multiplicative gap between our achievable load $L_A(\mathbf{m},\mathbf{w}_{\textrm{Com}}(\mathbf{m}))$ and the minimum load $L^*(\mathbf{m},\mathbf{w}_{\textrm{Com}}(\mathbf{m}))$, using computation-aware function assignment $\mathbf{w}_{\textrm{Com}}(\mathbf{m})$, is given in the following corollary, whose proof is in Appendix C.

\begin{corollary}\label{coro gap propotinal function assignment}
For a heterogeneous MapReduce computing system with $K$ nodes, computation load $\mathbf{m}=[m_1,\ldots,m_K]$, and function assignment $\mathbf{w}_{\textrm{Com}}(\mathbf{m})$, the multiplicative gap between our achievable communication load $L_A(\mathbf{m},\mathbf{w}_{\textrm{Com}}(\mathbf{m}))$ and the minimum load $L^*(\mathbf{m},\mathbf{w}_{\textrm{Com}}(\mathbf{m}))$ is within $16+70e$.
\end{corollary}

Corollary \ref{coro gap propotinal function assignment} implies that $\frac{L_A(\mathbf{m},\mathbf{w}_{\textrm{Com}}(\mathbf{m}))}{L_{Lower}(\mathbf{m},\mathbf{w}_{\textrm{Com}}(\mathbf{m}))}\le16+70e$. Thus, to obtain the multiplicative gap between $L_A(\mathbf{m},\mathbf{w}_{\textrm{Com}}(\mathbf{m}))$ and $L_{\textrm{Hom}}^*(\bar{m})$, we only need to obtain an upper bound of $\frac{L_{Lower}(\mathbf{m},\mathbf{w}_{\textrm{Com}}(\mathbf{m}))}{L_{\textrm{Hom}}^*(\bar{m})}$, given by
\begin{subequations}\label{eqn gap hom <0.55 1}
\begin{align}
  \frac{L_{Lower}(\mathbf{m},\mathbf{w}_{\textrm{Com}}(\mathbf{m}))}{L_{\textrm{Hom}}^*(\bar{m})}\le&\frac{\max_{\mathcal{T}\subset[K]}(1-\sum_{k\in\mathcal{T}}m_k)\sum_{k\in\mathcal{T}}w_k}{\frac{1-\bar{m}}{K\bar{m}}}\label{eqn gap hom <0.55 1 1}\\
  =&\frac{\max_{\mathcal{T}\subset[K]}(1-\sum_{k\in\mathcal{T}}m_k)\sum_{k\in\mathcal{T}}\frac{m_k}{\sum_{k\in[K]m_k}}}{\frac{1-\bar{m}}{\sum_{k\in[K]}m_k}}\notag\\
  =&\frac{\max_{\mathcal{T}\subset[K]}(1-\sum_{k\in\mathcal{T}}m_k)\sum_{k\in\mathcal{T}}m_k}{1-\bar{m}}\notag\\
  \le&\frac{1/4}{1-\bar{m}}\label{eqn gap hom <0.55 1 2}\\
  <&\frac{5}{9}.\notag
\end{align}
\end{subequations}
Here, \eqref{eqn gap hom <0.55 1 1} comes from the inequality $L^*_{\textrm{Hom}}(\bar{m})\ge \frac{1-\bar{m}}{K\bar{m}}$, and \eqref{eqn gap hom <0.55 1 2} comes from the inequality of arithmetic and geometric means. Thus, the multiplicative gap between $L_A(\mathbf{m},\mathbf{w}_{\textrm{Com}}(\mathbf{m}))$ and $L_{\textrm{Hom}}^*(\bar{m})$ is upper bounded by
\begin{align}
  \frac{L_A(\mathbf{m},\mathbf{w}_{\textrm{Com}}(\mathbf{m}))}{L_{\textrm{Hom}}^*(\bar{m})}=\frac{L_A(\mathbf{m},\mathbf{w}_{\textrm{Com}}(\mathbf{m}))}{L_{Lower}(\mathbf{m},\mathbf{w}_{\textrm{Com}}(\mathbf{m}))}\cdot \frac{L_{Lower}(\mathbf{m},\mathbf{w}_{\textrm{Com}}(\mathbf{m}))}{L_{\textrm{Hom}}^*(\bar{m})}<(16+70e)\cdot\frac{5}{9}<115.\notag
\end{align}
Note that when $\bar{m}\ge0.55$, $\frac{L_A(\mathbf{m},\mathbf{w}_{\textrm{Com}}(\mathbf{m}))}{L_{\textrm{Hom}}^*(\bar{m})}$ cannot be upper bounded by a constant for arbitrary $\mathbf{m}$. This is because when $\bar{m}\rightarrow1$, $L_{\textrm{Hom}}^*(\bar{m})$ approaches 0, but there always exist some $\mathbf{m}$ such that $\max_{\mathcal{T}\subset[K]}(1-\sum_{k\in\mathcal{T}}m_k)\sum_{k\in\mathcal{T}}m_k$ is close to $\frac{1}{4}$, which implies that the gap $\frac{L_{Lower}(\mathbf{m},\mathbf{w}_{\textrm{Com}}(\mathbf{m}))}{L_{\textrm{Hom}}^*(\bar{m})}\rightarrow\infty$.

\subsection{$\bar{m}\ge0.55$}
When $\bar{m}\ge0.55$, we use the Shuffle-aware function assignment, and the achievable load in Theorem \ref{thm achievable shuffle aware} is upper bounded by
\begin{subequations}\label{eqn gap hom >0.55 1}
\begin{align}
  L_A(\mathbf{m},\mathbf{w}_{\textrm{Shu}}(\mathbf{m}))\le&\frac{1}{\sum_{k=r+1}^{K}\frac{P_k}{1-P_k}}\notag\\
  =&\frac{1}{\sum_{k=r+1}^K\left(\frac{1}{1-P_k}-1\right)}\notag\\
  =&\frac{1}{\sum_{k=r+1}^K\frac{1}{1-P_k}-(K-r)}\notag\\
  =&\frac{1}{\left(1-\frac{1-\xi}{K-r}\right)\sum_{k=r+1}^K\frac{1}{1-m_k}-K+r}\notag\\
  \le&\frac{1}{\left(1-\frac{1-\xi}{K-r}\right)\frac{(K-r)^2}{\sum_{k=r+1}^K(1-m_k)}-K+r}\label{eqn gap hom >0.55 1 1}\\
  =&\frac{1}{\left(1-\frac{1-\xi}{K-r}\right)\frac{(K-r)^2}{(K-r)-\sum_{k=r+1}^Km_k}-K+r}\notag\\
  =&\frac{(K-r)-\sum_{k=r+1}^Km_k}{\left(1-\frac{1-\xi}{K-r}\right)(K-r)^2-(K-r)\left((K-r)-\sum_{k=r+1}^Km_k\right)}\notag\\
  =&\frac{(K-r)-\sum_{k=r+1}^Km_k}{(K-r)\left(K-r-1+\xi-(K-r)+\sum_{k=r+1}^Km_k\right)}\notag\\
  =&\frac{(K-r)-\sum_{k=r+1}^Km_k}{(K-r)\left(\sum_{k=1}^Km_k-1\right)}\label{eqn gap hom >0.55 1 2}
\end{align}
\end{subequations}
where \eqref{eqn gap hom >0.55 1 1} comes from the inequality of arithmetic and harmonic means. Thus, the multiplicative gap between $L_A(\mathbf{m},\mathbf{w}_{\textrm{Shu}}(\mathbf{m}))$ and $L_{\textrm{Hom}}^*(\bar{m})$ is given by
\begin{align}
\frac{L_A(\mathbf{m},\mathbf{w}_{\textrm{Shu}}(\mathbf{m}))}{L_{\textrm{Hom}}^*(\bar{m})}\le&\frac{\frac{(K-r)-\sum_{k=r+1}^Km_k}{(K-r)\left(\sum_{k=1}^Km_k-1\right)}}{\frac{1-\bar{m}}{\sum_{k=1}^Km_k}}\notag\\
=&\frac{\left((K-r)-\sum_{k=r+1}^Km_k\right)\sum_{k=1}^Km_k}{(K-r)\left(\sum_{k=1}^Km_k-1\right)(1-\frac{\sum_{k=1}^Km_k}{K})}\notag\\
<&\frac{K\left(K-\sum_{k=1}^Km_k\right)\sum_{k=1}^Km_k}{(K-r)\left(\sum_{k=1}^Km_k-1\right)(K-\sum_{k=1}^Km_k)}\notag\\
=&\frac{K\sum_{k=1}^Km_k}{(K-r)\left(\sum_{k=1}^Km_k-1\right)}.\label{eqn gap hom >0.55 2}
\end{align}
Since $\sum_{k=1}^Km_k=\xi+\sum_{k=r+1}^Km_k=\bar{m}K\ge0.55K$, we have
\begin{align}
  \xi\ge0.55K-\sum_{k=r+1}^Km_k>0.55K-(K-r)=r-0.45K.\label{eqn gap hom >0.55 3}
\end{align}
Using the definition of $r$ in \eqref{eqn general define r}, we have $\xi\le rm_r\le \frac{r}{K-r+1}$. Now, we use contradiction to proof $r<K$. If $r=K$, then we have $l_k=m_k$ for $k\in[K]$, which implies $\sum_{k\in[K]}m_k=1$. The average computation load is $\bar{m}=\frac{1}{K}\le\frac{1}{2}$, which is contradict to the assumption that $\bar{m}\ge0.55$. Thus, we proved $r<K$, and $\xi$ can be further upper bounded by $\xi\le \frac{r}{K-r+1}\le\frac{r}{2}$. Combining \eqref{eqn gap hom >0.55 3}, we have
$r-0.45K<\frac{r}{2}$, which implies $r<0.9K$. Then, \eqref{eqn gap hom >0.55 2} can be further upper bounded by
\begin{align}
  \frac{L_A(\mathbf{m},\mathbf{w}_{\textrm{Shu}}(\mathbf{m}))}{L_{\textrm{Hom}}^*(\bar{m})}<&\frac{K\sum_{k=1}^Km_k}{(K-r)\left(\sum_{k=1}^Km_k-1\right)}\notag\\
  &<\frac{K}{K-0.9K}\left(1+\frac{1}{\sum_{k=1}^Km_k-1}\right)\notag\\
  &\le10\left(1+\frac{1}{2\bar{m}-1}\right)<115.\notag
\end{align}
Thus, Corollary \ref{coro gap aware function assignment} is proved.

\section*{Appendix B: Lower Bound (Proof of Lemma \ref{thm converse load arbitrary})}
For some $\mathcal{W}\subseteq\{\phi_1,\ldots,\phi_Q\}$ and $\mathcal{M}\subseteq\{f_1,\ldots,f_N\}$, define $\mathcal{V}_{\mathcal{W},\mathcal{M}}\triangleq\{v_{q,n}:\phi_q\in\mathcal{W},f_n\in\mathcal{M}\}$. The proof is based on the following cut-set argument. Consider an arbitrary node set $\mathcal{T}$. For each node $k\in\mathcal{T}$, given its locally computed IVs $\mathcal{V}_{:,\mathcal{M}_k}$ and the communicated messages $\{X_1,\ldots,X_K\}$, it can successfully obtain its needed IVs $\mathcal{V}_{\mathcal{W}_k,:}$, where we use ``$:$'' to define the set of all possible indices. Thus, we have
\begin{align}
  H(\mathcal{V}_{\bigcup_{k\in\mathcal{T}}\mathcal{W}_k,:}|\mathcal{V}_{:,\bigcup_{k\in\mathcal{T}}\mathcal{M}_k},X_1,\ldots,X_K)=0.\label{eqn converse 1}
\end{align}
We also have
\begin{subequations}\label{eqn converse 2}
\begin{align}
  &H(\mathcal{V}_{\bigcup_{k\in\mathcal{T}}\mathcal{W}_k,:}|\mathcal{V}_{:,\bigcup_{k\in\mathcal{T}}\mathcal{M}_k})\notag\\
  =&H(\mathcal{V}_{\bigcup_{k\in\mathcal{T}}\mathcal{W}_k,\bigcup_{k\in\mathcal{T}}\mathcal{M}_k},\mathcal{V}_{\bigcup_{k\in\mathcal{T}}\mathcal{W}_k,\{f_1,\ldots,f_N\}\setminus\bigcup_{k\in\mathcal{T}}\mathcal{M}_k}|\mathcal{V}_{:,\bigcup_{k\in\mathcal{T}}\mathcal{M}_k})\notag\\
  =&H(\mathcal{V}_{\bigcup_{k\in\mathcal{T}}\mathcal{W}_k,\{f_1,\ldots,f_N\}\setminus\bigcup_{k\in\mathcal{T}}\mathcal{M}_k}|\mathcal{V}_{:,\bigcup_{k\in\mathcal{T}}\mathcal{M}_k})\notag\\
  =&H(\mathcal{V}_{\bigcup_{k\in\mathcal{T}}\mathcal{W}_k,\{f_1,\ldots,f_N\}\setminus\bigcup_{k\in\mathcal{T}}\mathcal{M}_k})\label{eqn converse 2 1}\\
  \ge&\sum_{k\in\mathcal{T}}w_kQ(1-\sum_{k\in\mathcal{T}}m_k)NT,\label{eqn converse 2 2}
\end{align}
\end{subequations}
where \eqref{eqn converse 2 1} comes from the fact that $\mathcal{V}_{\bigcup_{k\in\mathcal{T}}\mathcal{W}_k,\{f_1,\ldots,f_N\}\setminus\bigcup_{k\in\mathcal{T}}\mathcal{M}_k}$ are the IVs computed from files $\{f_1,\ldots,f_N\}\setminus\bigcup_{k\in\mathcal{T}}\mathcal{M}_k$, which are independent from $\mathcal{V}_{:,\bigcup_{k\in\mathcal{T}}\mathcal{M}_k}$; \eqref{eqn converse 2 2} comes from the inequality $|\bigcup_{k\in\mathcal{T}}\mathcal{M}_k|\le\sum_{k\in\mathcal{T}}m_kN$. Combining \eqref{eqn converse 1} and \eqref{eqn converse 2}, we have
\begin{align}
  \sum_{k\in\mathcal{T}}w_kQ(1-\sum_{k\in\mathcal{T}}m_k)NT\le& H(\mathcal{V}_{\bigcup_{k\in\mathcal{T}}\mathcal{W}_k,:}|\mathcal{V}_{:,\bigcup_{k\in\mathcal{T}}\mathcal{M}_k})-H(\mathcal{V}_{\bigcup_{k\in\mathcal{T}}\mathcal{W}_k,:}|\mathcal{V}_{:,\bigcup_{k\in\mathcal{T}}\mathcal{M}_k},X_1,\ldots,X_K)\notag\\
  =&I(\mathcal{V}_{\bigcup_{k\in\mathcal{T}}\mathcal{W}_k,:};X_1,\ldots,X_K|\mathcal{V}_{:,\bigcup_{k\in\mathcal{T}}\mathcal{M}_k})\notag\\
  =&H(X_1,\ldots,X_K|\mathcal{V}_{:,\bigcup_{k\in\mathcal{T}}\mathcal{M}_k})-H(X_1,\ldots,X_K|\mathcal{V}_{\bigcup_{k\in\mathcal{T}}\mathcal{W}_k,:},\mathcal{V}_{:,\bigcup_{k\in\mathcal{T}}\mathcal{M}_k})\notag\\
  \le&H(X_1,\ldots,X_K)=L^*(\mathbf{m},\mathbf{w})QNT\label{eqn converse 3}
\end{align}
Taking the maximum over $\mathcal{T}$ in \eqref{eqn converse 3}, we have
\begin{align}
  L^*(\mathbf{m},\mathbf{w})\ge\max\limits_{\mathcal{T}\subseteq[K]}\sum_{k\in\mathcal{T}}w_k(1-\sum_{k\in\mathcal{T}}m_k),\notag
\end{align}
and Lemma \ref{thm converse load arbitrary} is proved.

\setcounter{subsection}{0}
\section*{Appendix C: Proof of Corollary \ref{coro gap propotinal function assignment}}
Define $a\triangleq\frac{1}{\sum_{k\in[K]}m_k}$, then the achievable communication load in Theorem \ref{thm achievable computation aware} can be rewritten as
\begin{align}
  L_A(\mathbf{m},\mathbf{w}_{\textrm{Com}}(\mathbf{m}))\triangleq&\sum_{k=1}^{r}am_k(1-m_k)\!\notag\\
  &+\!\!\sum_{k=r+1}^{K}\!am_k\!\prod_{i=r+1}^{k}\!(1-P_i)\cdot\left[\xi+(K-k)\frac{1-\xi}{K-r}+\frac{1-\xi}{K-r}\sum_{i=r+1}^{k-1}\frac{1}{1-P_i}\right].\label{eqn achievable load computation aware with a}
\end{align}
Given the computation-aware function assignment \eqref{eqn computation aware assign}, the lower bound in Lemma \ref{thm converse load arbitrary} can be rewritten as
\begin{align}
L_{Lower}(\mathbf{m},\mathbf{w})\triangleq\max_{\mathcal{T}\subseteq[K]}(1-\sum_{k\in\mathcal{T}}m_k)a\sum_{k\in\mathcal{T}}m_k.\label{eqn converse computation aware}
\end{align}
We consider the following two cases to prove Corollary \ref{coro gap propotinal function assignment}: 1) $r=K$; 2) $r\le K-1$.

\subsection{$r=K$}
When $r=K$, the achievable load in \eqref{eqn achievable load computation aware with a} reduces to
\begin{align}
  L_A=\sum_{k=1}^{K}am_k(1-m_k).\label{eqn proof computation gap r=K 1}
\end{align}
We also have $\sum_{k\in[K]}m_k=1$. We consider two cases to prove the multiplicative gap: 1) $m_K\ge0.1$; 2) $m_K<0.1$.
\subsubsection{$m_K\ge0.1$}
By using the fact that $\sum_{k\in[K]}m_k=1$, the achievable load in \eqref{eqn proof computation gap r=K 1} can be upper bounded by
\begin{align}
  L_A\le a\left[\sum_{k=1}^{K-1}m_k+(1-m_K)m_K\right]=a(1-m_K)(1+m_K).\label{eqn proof computation gap r=K 2}
\end{align}
Letting $\mathcal{T}=\{K\}$ in \eqref{eqn converse computation aware}, we have
\begin{align}
L_{Lower}\ge(1-m_K)am_K.\label{eqn proof computation gap r=K 3}
\end{align}
By comparing \eqref{eqn proof computation gap r=K 2} and \eqref{eqn proof computation gap r=K 3}, the multiplicative gap is upper bounded by
\begin{align}
  \frac{L_A}{L^*}\le\frac{L_A}{L_{Lower}}\le\frac{1+m_K}{m_K}\le11.\notag
\end{align}

\subsubsection{$m_K<0.1$}
Recall that $m_1\le m_2\le \cdots\le m_K$. In this case, we have $m_k<0.1,\forall k\in[K]$. Define $\Omega(k)\triangleq\sum_{i=1}^km_i$. Then, we have $\Omega(1)=m_1<0.1$ and $\Omega(K)=1$. Since $m_k<0.1,\forall k\in[K]$, there must exist an integer $s$ such that $\Omega(s)=\sum_{k=1}^sm_i\in[0.45,0.55]$. By letting $\mathcal{T}=[s]$ in \eqref{eqn converse computation aware}, we have
\begin{align}
  L_{Lower}\ge(1-\sum_{k\in[s]}m_k)a\sum_{k\in[s]}m_k\ge 0.2475a.\label{eqn proof computation gap r=K 4}
\end{align}
The achievable load in \eqref{eqn proof computation gap r=K 1} is upper bounded by
\begin{align}
  L_A\le a\sum_{k=1}^Km_k=a.\label{eqn proof computation gap r=K 5}
\end{align}
By comparing \eqref{eqn proof computation gap r=K 4} and \eqref{eqn proof computation gap r=K 5}, the multiplicative gap is upper bounded by
\begin{align}
  \frac{L_A}{L^*}\le\frac{L_A}{L_{Lower}}\le\frac{a}{0.2475a}<5.\notag
\end{align}

Combining these two cases, the multiplicative gap when $r=K$ is upper bounded by 11.

\subsection{$r\le K-1$}
In \eqref{eqn achievable load computation aware with a}, define
\begin{align}
   L_{A,1}\triangleq&\sum_{k=1}^{r}am_k(1-m_k),\label{eqn proof computation gap r<K 1}\\
  L_{A,2}\triangleq&\sum_{k=r+1}^{K}\!am_k\!\prod_{i=r+1}^{k}\!(1-P_i)\cdot\left[\xi+(K-k)\frac{1-\xi}{K-r}+\frac{1-\xi}{K-r}\sum_{i=r+1}^{k-1}\frac{1}{1-P_i}\right].\label{eqn proof computation gap r<K 2}
\end{align}
Then, the multiplicative gap is given by
\begin{align}
  \frac{L_A}{L^*}\le\frac{L_{A,1}}{L_{Lower}}+\frac{L_{A,2}}{L_{Lower}}.\notag
\end{align}
In the following, we will first prove the multiplicative gap between $L_{A,1}$ and $L_{Lower}$, and then prove the multiplicative gap between $L_{A,2}$ and $L_{Lower}$.
\subsubsection{The multiplicative gap between $L_{A,1}$ and $L_{Lower}$}
We consider two cases to prove the gap: 1) $\sum_{k=1}^rm_k\le0.9$; 2) $\sum_{k=1}^rm_k>0.9$.

\textbf{Case 1} ($\sum_{k=1}^rm_k\le0.9$): $L_{A,1}$ is upper bounded by
\begin{align}
  L_{A,1}\le a\sum_{k=1}^{r}m_k.\label{eqn proof computation gap r<K g1 1}
\end{align}
Letting $\mathcal{T}=[r]$ in \eqref{eqn converse computation aware}, we have
\begin{align}
  L_{Lower}\ge(1-\sum_{k\in[r]}m_k)a\sum_{k\in[r]}m_k.\label{eqn proof computation gap r<K g1 2}
\end{align}
By comparing \eqref{eqn proof computation gap r<K g1 1} and \eqref{eqn proof computation gap r<K g1 2}, the multiplicative gap is upper bounded by
\begin{align}
  \frac{L_{A,1}}{L_{Lower}}\le\frac{1}{1-\sum_{k\in[r]}m_k}\le 10.\notag
\end{align}

\textbf{Case 2} ($\sum_{k=1}^rm_k>0.9$): Using the definition of $r$ in \eqref{eqn general define r}, we have
\begin{align}
  (K-r+1)m_r+\sum_{k=1}^{r-1}m_k\le 1,\label{eqn proof computation gap r<K g1 3}
\end{align}
which implies that $\sum_{k=1}^rm_k\le1-(K-r)m_r\le1$. Then, $L_{A,1}$ is upper bounded by
\begin{align}
  L_{A,1}\le a\sum_{k=1}^{r}m_k\le a.\label{eqn proof computation gap r<K g1 4}
\end{align}
We consider two sub-cases to prove the gap: 1) $m_r\ge0.1$; 2) $m_r<0.1$.
\begin{itemize}
  \item \textsf{Sub-case 1} ($m_r\ge0.1$): From \eqref{eqn proof computation gap r<K g1 3}, we have $m_r\le\frac{1}{K-r+1}\le\frac{1}{2}$. Letting $\mathcal{T}=\{r\}$ in \eqref{eqn converse computation aware}, we have
\begin{align}
  L_{Lower}\ge(1-m_r)am_r\ge0.09a.\label{eqn proof computation gap r<K g1 5}
\end{align}
By comparing \eqref{eqn proof computation gap r<K g1 4} and \eqref{eqn proof computation gap r<K g1 5}, the multiplicative gap is upper bounded by
\begin{align}
  \frac{L_{A,1}}{L_{Lower}}\le\frac{a}{0.09a}<12.\notag
\end{align}
  \item \textsf{Sub-case 2} ($m_r<0.1$): The proof is similar to the proof when $r=K$ and $m_K<0.1$. In this case, we have $m_k<0.1,\forall k\in[r]$.  Recall that $\Omega(k)=\sum_{i=1}^km_i$, then we have $\Omega(r)>0.9$ and $\Omega(1)<0.1$. Thus, there must exist an integer $s$ such that $\Omega(s)\in[0.45,0.55]$. By letting $\mathcal{T}=[s]$ in \eqref{eqn converse computation aware}, the lower bound is given by \eqref{eqn proof computation gap r=K 4}. Then, the multiplicative gap is upper bounded by
\begin{align}
  \frac{L_{A,1}}{L_{Lower}}\le\frac{a}{0.2475a}<5.\notag
\end{align}
\end{itemize}

By combining Case 1 and two sub-cases in Case 2, the multiplicative gap between $L_{A,1}$ and $L_{Lower}$ is upper bounded by 12.

\subsubsection{The multiplicative gap between $L_{A,2}$ and $L_{Lower}$}
We first consider the special case when $r=K-1$. In this case, $L_{A,2}$ is given by
\begin{align}
  L_{A,2}=am_K(1-P_K)\xi=am_K\xi\frac{1-m_K}{1-\frac{1-\xi}{K-(K-1)}}=am_K(1-m_K).\label{eqn proof computation gap r<K g2 1}
\end{align}
Letting $\mathcal{T}=\{K\}$ in \eqref{eqn converse computation aware}, we have
\begin{align}
  L_{Lower}\ge a(1-m_K)m_K.\label{eqn proof computation gap r<K g2 2}
\end{align}
By comparing \eqref{eqn proof computation gap r<K g2 1} and \eqref{eqn proof computation gap r<K g2 2}, the multiplicative gap is upper bounded by
\begin{align}
  \frac{L_{A,2}}{L_{Lower}}\le1.\notag
\end{align}

Now we consider the more general case $r\le K-2$. Recall that $m_{r+1}\le\cdots\le m_K$. We partition nodes $[r+1:K]$ into two disjoint subsets $[r+1:q]$ and $[q+1:K]$ such that $m_k<0.2,\forall k\in[r+1:q]$ and $m_k\ge0.2, \forall k\in[q+1:K]$\footnote{If $m_k<0.2,\forall k\in[r+1:K]$ or $m_k\ge0.2, \forall k\in[r+1:K]$, there will be only one set after the partition. Our proof is still applicable to these special cases.}. Then, $L_{A,2}$ can be rewritten as
\begin{align}
    L_{A,2}=&\sum_{k=r+1}^{q-1}\!am_k\!\prod_{i=r+1}^{k}\!(1-P_i)\cdot\left[\xi+(K-k)\frac{1-\xi}{K-r}+\frac{1-\xi}{K-r}\sum_{i=r+1}^{k-1}\frac{1}{1-P_i}\right]\notag\\
    &+am_q\!\prod_{i=r+1}^{q}\!(1-P_i)\cdot\left[\xi+(K-q)\frac{1-\xi}{K-r}\right]+am_q\!\prod_{i=r+1}^{q}\!(1-P_i)\cdot\frac{1-\xi}{K-r}\sum_{i=r+1}^{q-1}\frac{1}{1-P_i}\notag\\
    &+am_{q+1}\!\prod_{i=r+1}^{q+1}\!(1\!-\!P_i)\!\cdot\!\left[\xi\!+\!(K-q-1)\!\frac{1-\xi}{K-r}\right]+am_{q+1}\!\prod_{i=r+1}^{q+1}\!(1\!-\!P_i)\!\cdot\!\frac{1-\xi}{K-r}\sum_{i=r+1}^{q}\frac{1}{1\!-\!P_i}\notag\\
    &+\sum_{k=q+2}^{K}\!am_k\!\prod_{i=r+1}^{k}\!(1-P_i)\cdot\left[\xi+(K-k)\frac{1-\xi}{K-r}+\frac{1-\xi}{K-r}\sum_{i=r+1}^{k-1}\frac{1}{1-P_i}\right].\label{eqn proof computation gap r<K g2 3}
\end{align}
In \eqref{eqn proof computation gap r<K g2 3}, we define
\begin{align}
  L_{A,2}^1\triangleq&\sum_{k=r+1}^{q-1}\!am_k\!\prod_{i=r+1}^{k}\!(1-P_i)\cdot\left[\xi+(K-k)\frac{1-\xi}{K-r}+\frac{1-\xi}{K-r}\sum_{i=r+1}^{k-1}\frac{1}{1-P_i}\right]\notag\\
  &+am_q\!\prod_{i=r+1}^{q}\!(1-P_i)\cdot\frac{1-\xi}{K-r}\sum_{i=r+1}^{q-1}\frac{1}{1-P_i},\notag\\
  L_{A,2}^2\triangleq&\sum_{k=q+2}^{K}\!am_k\!\prod_{i=r+1}^{k}\!(1-P_i)\cdot\left[\xi+(K-k)\frac{1-\xi}{K-r}+\frac{1-\xi}{K-r}\sum_{i=r+1}^{k-1}\frac{1}{1-P_i}\right]\notag\\
  &+am_{q+1}\!\prod_{i=r+1}^{q+1}\!(1\!-\!P_i)\!\cdot\!\left[\xi\!+\!(K-q-1)\!\frac{1-\xi}{K-r}\right],\notag\\
  L_{A,2}^3\triangleq& am_q\!\prod_{i=r+1}^{q}\!(1-P_i)\cdot\left[\xi+(K-q)\frac{1-\xi}{K-r}\right],\notag\\
  L_{A,2}^4\triangleq& am_{q+1}\!\prod_{i=r+1}^{q+1}\!(1\!-\!P_i)\!\cdot\!\frac{1-\xi}{K-r}\sum_{i=r+1}^{q}\frac{1}{1\!-\!P_i}.\notag
\end{align}
We aim to compare $L_{A,2}^i$, for $i\in[4]$, to $L_{Lower}$ in \eqref{eqn converse computation aware} one by one, so as to obtain the multiplicative gap $\frac{L_{A,2}}{L_{Lower}}$.
\begin{lemma}\label{Lemma LA2 12}
When $r\le K-2$, the multiplicative gap between $L_{A,2}^1$ and $L_{Lower}$ in \eqref{eqn converse computation aware} is within $20e$, i.e., $\frac{L_{A,2}^1}{L_{Lower}}\le20e$, and the multiplicative gap between $L_{A,2}^2$ and $L_{Lower}$ in \eqref{eqn converse computation aware} is within $50e$, i.e., $\frac{L_{A,2}^1}{L_{Lower}}\le50e$.
\end{lemma}

The proof of Lemma \ref{Lemma LA2 12} is in Appendix D. Given Lemma \ref{Lemma LA2 12}, we only need to consider $L_{A,2}^3$ and $L_{A,2}^4$. Note that
\begin{align}
  1-P_k=\frac{1-m_k}{1-\frac{1-\xi}{K-r}}\le\frac{1-m_k}{1-\frac{1}{K-r}}\le\frac{1-m_k}{1-\frac{1}{2}}=2(1-m_k), \quad \forall k\in[r+1:K]\label{eqn proof computation gap r<K g2 4}
\end{align}
Then, $L_{A,2}^3$ is upper bounded by
\begin{align}
  L_{A,2}^3\le am_q\!\prod_{i=r+1}^{q}\!(1-P_i)\cdot\left[\xi+(K-r)\frac{1-\xi}{K-r}\right]\le am_q(1-P_q)\le 2am_q(1-m_q).\label{eqn proof computation gap r<K g2 5}
\end{align}
Letting $\mathcal{T}=\{q\}$ in \eqref{eqn converse computation aware}, we have
\begin{align}
  L_{Lower}\ge a(1-m_q)m_q.\label{eqn proof computation gap r<K g2 6}
\end{align}
Comparing \eqref{eqn proof computation gap r<K g2 5} and \eqref{eqn proof computation gap r<K g2 6}, we have
\begin{align}
  \frac{L_{A,2}^3}{L_{Lower}}\le2.\label{eqn proof computation gap r<K g2 6-1}
\end{align}
Similarly, $L_{A,2}^4$ is upper bounded by
\begin{align}
  L_{A,2}^4\le am_{q+1}\frac{1-\xi}{K-r}(q-r)(1-P_{q+1})\le 2am_{q+1}(1-m_{q+1}).\label{eqn proof computation gap r<K g2 7}
\end{align}
Letting $\mathcal{T}=\{q+1\}$ in \eqref{eqn converse computation aware}, we have
\begin{align}
  L_{Lower}\ge a(1-m_{q+1})m_{q+1}.\label{eqn proof computation gap r<K g2 8}
\end{align}
Comparing \eqref{eqn proof computation gap r<K g2 7} and \eqref{eqn proof computation gap r<K g2 8}, we have
\begin{align}
  \frac{L_{A,2}^4}{L_{Lower}}\le2.\label{eqn proof computation gap r<K g2 6-2}
\end{align}

Thus, when $r\le K-2$, combining Lemma \ref{Lemma LA2 12}, \eqref{eqn proof computation gap r<K g2 6-1}, and \eqref{eqn proof computation gap r<K g2 6-2}, we have
\begin{align}
  \frac{L_{A,2}}{L_{Lower}}=\frac{L_{A,2}^1}{L_{Lower}}+\frac{L_{A,2}^2}{L_{Lower}}+\frac{L_{A,2}^3}{L_{Lower}}+\frac{L_{A,2}^4}{L_{Lower}}\le 20e+50e+2+2=4+70e.\notag
\end{align}

Combining the special case when $r=K-1$ and the general case when $r\le K-2$, we also have
\begin{align}
  \frac{L_{A,2}}{L_{Lower}}\le4+70e.\notag
\end{align}
Thus, the multiplicative gap between the achievable load $L_{A}$ and the optimum $L^*$ when $r\le K-1$ is upper bounded by
\begin{align}
  \frac{L_A}{L^*}\le \frac{L_{A,1}}{L_{Lower}}+\frac{L_{A,2}}{L_{Lower}}\le12+4+70e=16+70e.\notag
\end{align}

Thus, Corollary \ref{coro gap propotinal function assignment} is proved.

\setcounter{subsection}{0}
\section*{Appendix D: Proof of Lemma \ref{Lemma LA2 12}}
\subsection{The multiplicative gap between $L_{A,2}^1$ and $L_{Lower}$}
First, $L_{A,2}^1$ is upper bounded by
\begin{align}
  L_{A,2}^1\le&\sum_{k=r+1}^{q-1}\!am_k\!\prod_{i=r+1}^{k}\!(1-P_i)\cdot\left[\xi+(K-k)\frac{1-\xi}{K-r}+\frac{1-\xi}{K-r}(k-r-1)\frac{1}{1-P_k}\right]\notag\\
  &+am_q\!\prod_{i=r+1}^{q-1}\!(1-P_i)\cdot\frac{1-\xi}{K-r}(q-r-1)\notag\\
  =&a\sum_{k=r+1}^{q-1}\prod_{i=r+1}^{k}\!(1-P_i)\cdot\left[\left(\xi+(K-k)\frac{1-\xi}{K-r}\right)m_k+\frac{1-\xi}{K-r}(k-r)m_{k+1}\right]\notag\\
  \le&a\sum_{k=r+1}^{q-1}\prod_{i=r+1}^{k}\!(1-P_i)\cdot m_{k+1}\left[\xi+(K-k)\frac{1-\xi}{K-r}+\frac{1-\xi}{K-r}(k-r)\right]\notag\\
  =&a\sum_{k=r+1}^{q-1}m_{k+1}\prod_{i=r+1}^{k}\!(1-P_i)\label{eqn AL21 1}
\end{align}
In \eqref{eqn AL21 1}, $\prod_{i=r+1}^{k}\!(1-P_i)$ is upper bounded by
\begin{align}
  \prod_{i=r+1}^{k}\!(1-P_i)=&\left(\frac{1}{1-\frac{1-\xi}{K-r}}\right)^{k-r}\prod_{i=r+1}^{k}\!(1-m_i)\notag\\
  \le&\left(\frac{1}{1-\frac{1-\xi}{K-r}}\right)^{K-r}\prod_{i=r+1}^{k}\!(1-m_i)\notag\\
  \le&\left(\frac{1}{1-\frac{1-\xi}{K-r}}\right)^{\frac{K-r}{1-\xi}}\prod_{i=r+1}^{k}\!(1-m_i)\notag\\
  =&\left(\frac{\frac{K-r}{1-\xi}}{\frac{K-r}{1-\xi}-1}\right)^{\frac{K-r}{1-\xi}}\prod_{i=r+1}^{k}\!(1-m_i)\notag\\
  =&\left(1+\frac{1}{\frac{K-r}{1-\xi}-1}\right)^{\frac{K-r}{1-\xi}}\prod_{i=r+1}^{k}\!(1-m_i)\notag\\
  =&\left(1+\frac{1}{\frac{K-r}{1-\xi}-1}\right)\cdot \left(1+\frac{1}{\frac{K-r}{1-\xi}-1}\right)^{\frac{K-r}{1-\xi}-1}\prod_{i=r+1}^{k}\!(1-m_i)\notag\\
  \le&2e\prod_{i=r+1}^{k}\!(1-m_i),\label{eqn AL21 2}
\end{align}
where the last inequality comes from the fact that $\frac{K-r}{1-\xi}-1\ge K-r-1\ge1$. Substituting \eqref{eqn AL21 2} into \eqref{eqn AL21 1}, $L_{A,2}^1$ is upper bounded by
\begin{align}
  L_{A,2}^1\le 2ea\sum_{k=r+1}^{q-1}m_{k+1} \prod_{i=r+1}^{k}\!(1-m_i).\label{eqn AL21 3}
\end{align}
We consider two cases to prove the gap: 1) $\sum_{k=r+1}^qm_k\le 0.9$; 2) $\sum_{k=r+1}^qm_k> 0.9$.

\subsubsection{$\sum_{k=r+1}^qm_k\le 0.9$}
Letting $\mathcal{T}=[r+1:q]$ in \eqref{eqn converse computation aware}, we have
\begin{align}
  L_{Lower}\ge a(1-\sum_{k=r+1}^qm_{k})\sum_{k=r+1}^qm_{k}.\label{eqn AL21 4}
\end{align}
Comparing \eqref{eqn AL21 3} and \eqref{eqn AL21 4}, we have
\begin{align}
  \frac{L_{A,2}^1}{L_{Lower}}\le&\frac{2ea\sum_{k=r+1}^{q-1}m_{k+1} }{a(1-\sum_{k=r+1}^qm_{k})\sum_{k=r+1}^qm_{k}}\notag\\
  \le&\frac{2ea\sum_{k=r+1}^{q}m_{k}}{a(1-\sum_{k=r+1}^qm_{k})\sum_{k=r+1}^qm_{k}}\notag\\
  =&\frac{2ea}{a(1-\sum_{k=r+1}^qm_{k})}\le20e.\label{eqn AL21 5}
\end{align}

\subsubsection{$\sum_{k=r+1}^qm_k> 0.9$}
Define $\Omega'(k)\triangleq\sum_{i=r+1}^km_i$. We have $\Omega'(r+1)<0.2$ and $\Omega'(q)>0.9$. Similar to the proof when $r=K$ and $m_K<0.1$, since $m_k<0.2,\forall k\in[r+1:q]$, there must exist an integer $s$ such that $\Omega'(s)\in[0.4:0.6]$. Letting $\mathcal{T}=[r+1:s]$ in \eqref{eqn converse computation aware}, we have
\begin{align}
  L_{Lower}\ge a(1-\sum_{k=r+1}^sm_{k})\sum_{k=r+1}^sm_{k}\ge0.24a.\label{eqn AL21 6}
\end{align}
Since $\frac{1-m_k}{1-0.2}>1,\forall k\in[r+1:q]$, $L_{A,2}^1$ in \eqref{eqn AL21 3} is upper bounded by
\begin{align}
  L_{A,2}^1\le \frac{2ea}{0.8}\sum_{k=r+1}^{q-1}m_{k+1} \prod_{i=r+1}^{k+1}\!(1-m_i).\label{eqn AL21 7}
\end{align}
Now, we use induction to prove $\sum_{k=r+1}^{q-1}m_{k+1} \prod_{i=r+1}^{k+1}\!(1-m_i)\le(1-m_{r+1})$. If
\begin{align}
  \sum_{k=u}^{q-1}m_{k+1} \prod_{i=r+1}^{k+1}\!(1-m_i)\le\prod_{i=r+1}^{u}(1-m_i),\notag
\end{align}
then we have
\begin{align}
  \sum_{k=u-1}^{q-1}m_{k+1} \prod_{i=r+1}^{k+1}\!(1-m_i)\le&\prod_{i=r+1}^{u}(1-m_i)+m_u\prod_{i=r+1}^{u}(1-m_i)\notag\\
  =&(1+m_u)\prod_{i=r+1}^{u}(1-m_i)\notag\\
  \le&\prod_{i=r+1}^{u-1}(1-m_i).\notag
\end{align}
Letting $u=q-1$, we have
\begin{align}
   m_{q}\prod_{i=r+1}^{q}(1-m_i)=m_{q}(1-m_q)\prod_{i=r+1}^{q-1}(1-m_i)\le\prod_{i=r+1}^{q-1}(1-m_i),\notag
 \end{align}
which is true. Thus, letting $u=r+1$, we prove that
\begin{align}
  \sum_{k=r+1}^{q-1}m_{k+1} \prod_{i=r+1}^{k+1}\!(1-m_i)\le(1-m_{r+1}).\label{eqn AL21 8}
\end{align}
Thus, using \eqref{eqn AL21 6}, \eqref{eqn AL21 7}, and \eqref{eqn AL21 8}, the multiplicative gap is upper bounded by
\begin{align}
  \frac{L_{A,2}^1}{L_{Lower}}\le\frac{\frac{2ea}{0.8}\sum_{k=r+1}^{q-1}m_{k+1} \prod_{i=r+1}^{k+1}\!(1-m_i)}{0.24a}\le\frac{\frac{2ea}{0.8}(1-m_{r+1})}{0.24a}\le12e.\notag
\end{align}

Combining the two cases, we prove that $\frac{L_{A,2}^1}{L_{Lower}}\le20e$.

\subsection{The Multiplicative gap between $L_{A,2}^2$ and $L_{Lower}$}
$L_{A,2}^2$ is upper bounded by
\begin{subequations}\label{eqn AL22 1}
  \begin{align}
  L_{A,2}^2\le&\sum_{k=q+2}^{K}\!am_k\!\prod_{i=r+1}^{k}\!(1-P_i)\cdot\left[\xi+(K-k)\frac{1-\xi}{K-r}+\frac{1-\xi}{K-r}(k-r-1)\frac{1}{1-P_k}\right]\notag\\
  &+am_{q+1}\!\prod_{i=r+1}^{q+1}\!(1-P_i)\cdot\left[\xi+(K-q-1)\frac{1-\xi}{K-r}\right]\notag\\
  \le&a\sum_{k=q+1}^{K}\prod_{i=r+1}^{k}\!(1-P_i)\cdot\left[\left(\xi+(K-k)\frac{1-\xi}{K-r}\right)m_k+\frac{1-\xi}{K-r}(k-r)m_{k+1}\right]\label{eqn AL22 1 1}\\
  \le&a\sum_{k=q+1}^{K}\prod_{i=r+1}^{k}\!(1-P_i)\label{eqn AL22 1 2}\\
  \le&2ea\sum_{k=q+1}^{K}\prod_{i=r+1}^{k}\!(1-m_i)\label{eqn AL22 1 3}\\
  \le&2ea\sum_{k=q+1}^{K}\prod_{i=q+1}^{k}\!(1-m_i)\notag\\
  \le&2ea\sum_{k=q+1}^{K}(1-m_{q+1})^{k-q}\notag\\
  =&2ea(1-m_{q+1})\frac{1-(1-m_{q+1})^{K-q}}{m_{q+1}}\notag\\
  \le&2ea(1-m_{q+1})\frac{1}{m_{q+1}},\label{eqn AL22 1 4}
\end{align}
\end{subequations}
where we define $m_{K+1}\triangleq1$ in \eqref{eqn AL22 1 1}; \eqref{eqn AL22 1 2} comes from the fact that $m_k\le 1,\forall k\in[r+1:K+1]$; \eqref{eqn AL22 1 3} comes from \eqref{eqn AL21 2}. Letting $\mathcal{T}=\{q+1\}$ in \eqref{eqn converse computation aware}, we have
\begin{align}
  L_{Lower}\ge a(1-m_{q+1})m_{q+1}.\label{eqn AL22 1 4}
\end{align}
Then, the multiplicative gap is upper bounded by
\begin{align}
  \frac{L_{A,2}^2}{L_{Lower}}\le\frac{2ea(1-m_{q+1})\frac{1}{m_{q+1}}}{a(1-m_{q+1})m_{q+1}}=2e\frac{1}{m_{q+1}^2}\le50e.\notag
\end{align}

Thus, Lemma \ref{Lemma LA2 12} is proved.

\bibliographystyle{IEEEtran}
\bibliography{IEEEabrv,journal}

\begin{thebibliography}{10}
\providecommand{\url}[1]{#1}
\csname url@samestyle\endcsname
\providecommand{\newblock}{\relax}
\providecommand{\bibinfo}[2]{#2}
\providecommand{\BIBentrySTDinterwordspacing}{\spaceskip=0pt\relax}
\providecommand{\BIBentryALTinterwordstretchfactor}{4}
\providecommand{\BIBentryALTinterwordspacing}{\spaceskip=\fontdimen2\font plus
\BIBentryALTinterwordstretchfactor\fontdimen3\font minus
  \fontdimen4\font\relax}
\providecommand{\BIBforeignlanguage}[2]{{%
\expandafter\ifx\csname l@#1\endcsname\relax
\typeout{** WARNING: IEEEtran.bst: No hyphenation pattern has been}%
\typeout{** loaded for the language `#1'. Using the pattern for}%
\typeout{** the default language instead.}%
\else
\language=\csname l@#1\endcsname
\fi
#2}}
\providecommand{\BIBdecl}{\relax}
\BIBdecl

\bibitem{MapReduceOrigin}
J.~Dean and S.~Ghemawat, ``Map{R}educe: Simplified data processing on large
  clusters,'' \emph{Commun. ACM}, vol.~51, no.~1, pp. 107--113, Jan. 2008.

\bibitem{EC2}
Z.~{Zhang}, L.~{Cherkasova}, and B.~T. {Loo}, ``Performance modeling of
  mapreduce jobs in heterogeneous cloud environments,'' in \emph{IEEE 6th Int.
  Conf. Cloud Comput. ({CLOUD})}, June 2013, pp. 839--846.

\bibitem{LiSongZeMapReduce}
S.~{Li}, M.~A. {Maddah-Ali}, Q.~{Yu}, and A.~S. {Avestimehr}, ``A fundamental
  tradeoff between computation and communication in distributed computing,''
  \emph{IEEE Trans. Inf. Theory}, vol.~64, no.~1, pp. 109--128, Jan 2018.

\bibitem{JimingyueNewCombinatorial}
N.~{Woolsey}, R.~{Chen}, and M.~{Ji}, ``A new combinatorial design of coded
  distributed computing,'' in \emph{IEEE ISIT}, June 2018.

\bibitem{LeveragingCoding}
K.~{Konstantinidis} and A.~{Ramamoorthy}, ``Leveraging coding techniques for
  speeding up distributed computing,'' in \emph{IEEE GLOBECOM}, Dec 2018.

\bibitem{Yanqifa3C}
Q.~{Yan}, S.~{Yang}, and M.~{Wigger}, ``Storage, computation, and
  communication: A fundamental tradeoff in distributed computing,'' in
  \emph{IEEE ITW}, Nov 2018.

\bibitem{Loadscheduling}
\BIBentryALTinterwordspacing
M.~Zhao, W.~Wang, Y.~Wang, and Z.~Zhang, ``Load scheduling for distributed edge
  computing: A communication-computation tradeoff,'' \emph{Peer-to-Peer
  Networking and Applications}, Oct 2018. [Online]. Available:
  \url{https://doi.org/10.1007/s12083-018-0695-4}
\BIBentrySTDinterwordspacing

\bibitem{LisongzeScalable}
S.~{Li}, Q.~{Yu}, M.~A. {Maddah-Ali}, and A.~S. {Avestimehr}, ``A scalable
  framework for wireless distributed computing,'' \emph{IEEE/ACM Trans.
  Networking}, vol.~25, no.~5, pp. 2643--2654, Oct 2017.

\bibitem{ChenjinyuanWirelessMR}
F.~{Li}, J.~{Chen}, and Z.~{Wang}, ``Wireless {M}ap{R}educe distributed
  computing,'' in \emph{IEEE ISIT}, June 2018.

\bibitem{Wangchengwei}
M.~{Kiamari}, C.~{Wang}, and A.~S. {Avestimehr}, ``On heterogeneous coded
  distributed computing,'' in \emph{IEEE GLOBECOM}, Dec 2017, pp. 1--7.

\bibitem{edge-facilitated}
------, ``Coding for edge-facilitated wireless distributed computing with
  heterogeneous users,'' in \emph{51st Asilomar Conf. Signals, Syst. Comput.},
  Oct 2017, pp. 536--540.

\bibitem{Chenjinyuan}
N.~{Shakya}, F.~{Li}, and J.~{Chen}, ``On distributed computing with
  heterogeneous communication constraints,'' in \emph{52nd Asilomar Conf.
  Signals, Syst. Comput.}, Oct 2018, pp. 1795--1799.

\bibitem{JimingyueCascade}
N.~Woolsey, R.~Chen, and M.~Ji, ``Cascaded coded distributed computing on
  heterogeneous networks,'' in \emph{IEEE ISIT}, July 2019.

\bibitem{JimingyueNoncascade}
\BIBentryALTinterwordspacing
------, ``Coded distributed computing with heterogeneous function
  assignments,'' 2019. [Online]. Available:
  \url{http://arxiv.org/abs/1902.10738}
\BIBentrySTDinterwordspacing

\bibitem{decentralized}
M.~A. Maddah-Ali and U.~Niesen, ``Decentralized coded caching attains
  order-optimal memory-rate tradeoff,'' \emph{IEEE/ACM Trans. Networking},
  vol.~23, no.~4, pp. 1029--1040, Aug 2015.

\bibitem{Wangsinong}
\BIBentryALTinterwordspacing
S.~Wang, W.~Li, X.~Tian, and H.~Liu, ``Fundamental limits of heterogenous
  cache,'' 2015. [Online]. Available: \url{http://arxiv.org/abs/1504.01123}
\BIBentrySTDinterwordspacing

\end{thebibliography}
\end{document}